\documentclass[aps,pra,reprint,superscriptaddress]{revtex4-2}
\usepackage[utf8]{inputenc}
\usepackage{amsmath}			% include some special mathematical symbols
\usepackage{amssymb}			% include some special mathematical symbols
\usepackage{graphicx}			% for graphic placement	
\usepackage{xcolor}				% for special text coloring
\usepackage{physics}            % commands for physics equations
\usepackage{siunitx}			% proper display of physical units
\usepackage{tikz}
\usetikzlibrary{calc, decorations.markings, decorations.pathmorphing, decorations.pathreplacing, decorations.fractals, arrows.meta,calendar,intersections,mindmap,folding,patterns,plothandlers,plotmarks,shapes,shadings,shadows,topaths,trees}
\usepackage{pgf}
\usepackage{ulem}
\pagecolor{white}
\usepackage{comment}
\usepackage{pdfpages} % include pdfs
\usepackage{pgffor} % for loops
\usepackage{xr} % referencing the supplement

\makeatletter
\AtBeginDocument{\let\LS@rot\@undefined}
\makeatother

\def\supplementfilename{SI}

\pdfximage{\supplementfilename.pdf}
\def\numbersupplementpages{\the\pdflastximagepages}

\pdfximage{\supplementfilename.pdf}
\def\numbersupplementpages{\the\pdflastximagepages}
\newcommand{\ie}{i.\,e.}
\newcommand{\eg}{e.\,g.}

\newcommand{\eqspace}{\:}	% additional space between equation end and comma/full stop
\newcommand{\sref}[1]{section~\ref{#1}}

\newcommand{\fref}[1]{figure~\ref{#1}}
\newcommand{\Fref}[1]{Figure~\ref{#1}}
\newcommand{\tref}[1]{table~\ref{#1}}

\newcommand{\sieqref}[1]{(\textcolor{blue}{S-#1})}
\definecolor{sicolor}{rgb}{0,0,0}
\newcommand{\sisref}[1]{section~\textcolor{sicolor}{S-#1}}

\newcommand{\sifref}[1]{figure~\textcolor{sicolor}{S-#1}}

\newcommand{\defeq}{\mathrel{\vcenter{%
			\baselineskip0.5ex\lineskiplimit0pt\hbox{%
				\scriptsize.}\hbox{\scriptsize.}}} =}					% nice colon-equals symbol
\newcommand{\half}{\tfrac{1}{2}}							% small fraction 1/2
\newcommand{\transpose}{^\intercal}							% transposition
\newcommand{\inverse}{^{-1}}								% inverse as exponent
\newcommand{\radial}{r}										% radial position
\newcommand{\channelradius}{R}								% channel radius
\newcommand{\velocity}{u}									% velocity
\newcommand{\dynvisc}{\eta}									% dynamic viscosity
\newcommand{\straintensor}{\dot{S}}                         % Rate of strain tensor
\newcommand{\rateofstrain}{|\straintensor|}					% rate of strain
\newcommand{\elongrate}{\dot{\varepsilon}}                  % elongation rate
\newcommand{\stress}{\sigma}								% stress
\newcommand{\pgrad}{G}										% constant pressure gradient
\newcommand{\flowrate}{\Omega}								% volume flux
\newcommand{\kronecker}[1]{\delta_{#1}}						% Kronecker symbol
\newcommand{\capillarynumber}{\mathrm{Ca}}                  % Capillary number
\newcommand{\shearmodulus}{\mu}                             % Shear modulus
\newcommand{\bulkmodulus}{\kappa}                           % bulk modulus
\newcommand{\poissonratio}{\nu}                             % Poisson ratio
\newcommand{\youngsmodulus}{E}                              % Young's modulus
\newcommand{\vonmises}{\sigma_\mathrm{vM}}                  % Von Mises effective stress
\newcommand{\fluidstress}{\sigma_\mathrm{f}}                % Fluid stress
\newcommand{\cellradius}{R_\mathrm{c}}						% Cell radius
\newcommand{\avgelongstress}{\bar{\sigma}_\mathrm{f}^\mathrm{elong}}
\newcommand{\SG}[1]{{\color{red}SG:~#1}}

\usepackage{hyperref}
\hypersetup{
	pdftitle={Predicting cell stress and strain during extrusion 3D bioprinting},    % title
	pdfauthor={S. J. M\"uller and S. Gekle},     % authors
	colorlinks=true,		% false: boxed links; true: colored links
	linkcolor=blue,			% color of internal links (change box color with linkbordercolor)
	citecolor=blue,			% color of links to bibliography
	filecolor=magenta,		% color of file links
	urlcolor=cyan			% color of external links
}
\begin{document}
%\preprint{}
\title{Predicting cell stress and strain during extrusion bioprinting}
\author{Sebastian J. M\"uller}
\affiliation{Biofluid Simulation and Modeling, Theoretische Physik VI, Universität Bayreuth, 95440 Bayreuth, Germany}
\author{Ben Fabry}
\affiliation{Department of Physics, Friedrich-Alexander University Erlangen-Nürnberg, 91054 Erlangen, Germany}
\author{Stephan Gekle}
\affiliation{Biofluid Simulation and Modeling, Theoretische Physik VI, Universität Bayreuth, 95440 Bayreuth, Germany}
\homepage[]{www.gekle.physik.uni-bayreuth.de}
%\thanks{}
\date{\today}
\begin{abstract}

Bioprinting of living cells can cause major shape deformations, which may severely affect cell survival and functionality.
While the shear stresses occurring during cell flow through the printer nozzle have been quantified to some extent, the extensional stresses occurring as cells leave the nozzle into the free printing strand have been mostly ignored.
Here we use Lattice-Boltzmann simulations together with a finite-element based cell model to study cell deformation at the nozzle exit.
Our simulation results are in good qualitative agreement with experimental microscopy images.
We show that for cells flowing in the center of the nozzle extensional stresses can be significant, while for cells flowing off-center their deformation is dominated by the shear flow inside the nozzle.
From the results of these simulations, we develop two simple methods that only require the printing parameters (nozzle diameter, flow rate, bioink rheology) to (i)~accurately predict the maximum cell stress occurring during the 3D bioprinting process and (ii)~approximately predict the cell strains caused by the elongational flow at the nozzle exit.
%\SG{ General remark: it would be nice, if you could make some 3-4 movies for different situations and include them as further SI material }\SJM{$\checkmark$}

%In 3D bioprinting, the survival and maintenance of functionality of the printed cells is crucial for the production of functioning biological tissues.
%While cell damage due to biochemical mechanisms can --- in principle --- be eliminated by a proper material composition, the physical stresses occurring during the extrusion process inevitably act on the cell.
%\\
%These stresses arise from the coupling between the fluid motion of the bioink, which is determined by the printing parameters and the bioink rheology on the one hand, and the mechano-elastic response of the cell on the other.
%We use three dimensional simulations of a hyperelastic cell in shear thinning bioinks to investigate the cell stress and strains as function of the printing parameters and the bioink rheology during its flow through a bioprinter nozzle and the transition into the free bioink strand.
%Since cells migrate towards the channel center inside a capillary, we consider the scenario of centered cells passing the nozzle in particular.
%From the results of these simulations, we develop two simple methods that only require the printing parameters to (i) accurately predict the maximum cell stress occurring during the 3D bioprinting process and (ii) approximate the cell strains caused by the elongational flow at the nozzle exit.
\end{abstract}
%\keywords{insert suggested keywords - APS authors don't need to do this}
-\maketitle
\section{Introduction}
\label{sec:introduction}
The aim of 3D bioprinting is to transfer the well-established techniques of conventional 3D printing to the fabrication of functional, living tissues.
The material to be printed typically consists of a chemically complex hydrogel, termed the bioink, in which living cells are suspended.
This technology promises to become a major breakthrough, e.g.\ for cancer studies or --~in the long run~-- organ replacements \cite{sun_bioprinting_2020,barrs_biomaterials_2020,levato_shape_2020,groll_biofabrication_2016,malda_25th_2013-1}.
A key obstacle, however, remains to ensure the survival and functionality of cells during and after the fabrication process.
Possible causes for cell damage are numerous, but can be broadly classified into insufficient bio-compatibility and mechanical damage.
%\BF{I would not call this "chemical damage". Generally, toxic reactions are rarely a problem. Rather, bioinks are not really bio-compatible. They are often bio-inert, but that is not always sufficient to ensure cell survival. Many cells need matrix adhesions and space to spread and grow for their suvival.}
The former arises from direct interaction between the cell and the surrounding bioink and has been intensively studied \cite{bock_tgf-1-modified_2018,esser_promoting_2019,hauptstein_hyaluronic_2020,mancini_composite_2020,mueller_effects_2021,roshanbinfar_electroconductive_2018,schmidt_differential_2020,weizel_complex_2020}.

Mechanical damage, by contrast, arises from hydrodynamic stresses as the cell passes from the reservoir through the printing nozzle, transitions into the printing strand, and finally comes to rest in the printed construct.
It has been shown that even after optimizing biological and chemical conditions~\cite{fischer_calcium_2022}, such hydrodynamic stresses remain a crucial source of cell damage and death \cite{han_study_2021,poologasundarampillai_real-time_2021,emmermacher_engineering_2020,boularaoui_overview_2020,ruther_biofabrication_2019, shi_shear_2018, paxton_proposal_2017, ouyang_effect_2016, blaeser_controlling_2015, snyder_mesenchymal_2015, zhao_influence_2015}.
Understanding these mechanical stress response processes is notoriously difficult as they result from an interplay between the complex rheology of the bioink, which is typically shear thinning \cite{hazur_improving_2020,hu_improving_2021,nadernezhad_rheological_2020,weis_evaluation_2018}, and the even more complex viscoelastic response of the cell itself \cite{muller_hyperelastic_2020-2, fregin_high-throughput_2019, saadat_immersed-finite-element_2018, mokbel_numerical_2017, mietke_extracting_2015, otto_real-time_2015, huber_emergent_2013, rodriguez_review_2013, gao_rheology_2011,kollmannsberger_linear_2011}. 
%The latter is typically assumed to be viscoelastic \cite{muller_hyperelastic_2020-2, fregin_high-throughput_2019, saadat_immersed-finite-element_2018, mokbel_numerical_2017, mietke_extracting_2015, otto_real-time_2015, gao_rheology_2011}.
Despite these difficulties, certain progress towards reliable theoretical estimates of the cell stress inside printing needles has been achieved~\cite{boularaoui_overview_2020,ning_characterization_2018,han_study_2021}.
As a starting point, the fluid shear stress profiles within printing nozzles have been computed theoretically \cite{emmermacher_engineering_2020,muller_flow_2020-1,paxton_proposal_2017}.
%3D bioprinter nozzles are commonly available either as syringe needles with a cylindrical geometry or as conically shaped pipetting tips.
Some experimental studies correlated such stress calculations with cell viability or functionality \cite{lemarie_rheology_2021,han_study_2021,fakhruddin_effects_2018,ning_characterization_2018,ouyang_effect_2016,billiet_3d_2014,bae_microfluidic_2016,nair_characterization_2009,chang_effects_2008}.
These studies, however, investigated hydrodynamic stresses only up to the end of the printing nozzle.
At the transition from the nozzle exit into the free strand, the flow profile transitions rapidly from a Poiseuille-like profile in the nozzle to a plug flow profile inside the free bioink strand.
This transition is accompanied by sizable radial flows whose effect on cell deformation and therefore cell damage has so far not been experimentally or theoretically quantified.

In this work, we start with fully three-dimensional Lattice Boltzmann calculations for the flow profile of shear thinning fluids \cite{muller_flow_2020-1} at the exit of a printing nozzle.
%To investigate cell stress and strains inside and during exit from the printing nozzle, we then employ our recently developed hyperelastic cell model \cite{muller_hyperelastic_2020-2} which includes explicit two-way coupling between bioink and cellular mechanics. 
To investigate cell stress and strains inside and during exit from the printing nozzle, we then employ our recently developed hyperelastic cell model \cite{muller_hyperelastic_2020-2} which includes explicit two-way coupling between bioink and cellular mechanics, and show its qualitative match with experimental micrographs taken during the printing process.
From these investigations, we finally develop simple methods to predict the cell stress and strains occurring during bioprinting processes, and specifically during nozzle exit, by only using the printing parameters, \ie, the nozzle diameter, the bioink rheology, and the volumetric flow rate, as input quantities.
For this, we combine the classical theories of Jeffery \cite{jeffery_motion_1922} and Roscoe \cite{roscoe_rheology_1967} with our semi-analytical flow computations of shear thinning bioinks in capillaries \cite{muller_flow_2020-1}.
\section{Methods and setup}
\label{sec:methods}
%In this section we introduce the numerical methods used in the present work for our flow calculations, flow simulations, our hyperelastic cell model, and the setups employed to simulate the 3D bioprinting process of a cell.
%

\subsection{Flow dynamics: Lattice-Boltzmann simulations}
\label{sec:methods-fluid}
In our simulations, we employ a fully three-dimensional fluid dynamics solver.
We use the implementation of the Lattice Boltzmann method \cite{kruger_lattice_2017} provided by the software package ESPResSo \cite{limbach_espressoextensible_2006,roehm_lattice_2012-2}, which we extended with algorithms to allow for the simulation of free-slip surfaces \cite{schlenk_parallel_2018-1} and shear thinning fluids \cite{muller_flow_2020-1}.
Using an immersed-boundary algorithm \cite{devendran_immersed_2012-1,bacher_clustering_2017-1,saadat_immersed-finite-element_2018,bacher_computational_2019-2}, we couple our cell model (\sref{sec:methods-cell}) to the flow.
%
%\subsection{Flow dynamics: Lattice-Boltzmann simulations}
%%\label{sec:methods-fluid}
%\SG{ I would describe only LBM and place this after ''Bioink rheology'' }
%We use two numerical methods for flow calculation: Our semi-analytical method to solve the Navier-Stokes equations in a cylindrical geometry and our fully three-dimensional flow solver coupled to the hyperelastic cell model for our simulations of the printing process.
%\\
%The first method, modeling the nozzle channel, considers the undisturbed flow, \ie, without the cell. 
%In our previous study, we introduced this method to be able to calculate the velocity, shear rate, viscosity, and shear stress, profiles for an inelastic shear thinning fluid in a cylindrical nozzle \cite{muller_flow_2020-1}.
%The central assumptions --- a laminar, uni-axial, pressure driven, flow --- are usually applicable for the description of bioink extrusion.
%\\
%The second method is our fully three-dimensional fluid dynamics solver.
%We use the implementation of the Lattice Boltzmann Method provided by the software package ESPResSo \cite{limbach_espressoextensible_2006,roehm_lattice_2012-2}, which we extended with algorithms to allow the simulation of free-slip surfaces \cite{schlenk_parallel_2018-1} and shear thinning fluids \cite{muller_flow_2020-1}.
%Using an immersed-boundary algorithm \cite{devendran_immersed_2012-1,bacher_clustering_2017-1,saadat_immersed-finite-element_2018,bacher_computational_2019-2}, we couple our cell model (\sref{sec:methods-cell}) to the flow.
%
\subsection{Bioink rheology}
\label{sec:methods-bioink-rheology}
\begin{figure}
	\centering
	\caption{\label{fig-viscosity-shearrate-materials}(a) Viscosity as function of the shear rate for the six different degrees of shear thinning ($\alpha=0$, $0.15$, $0.3$, $0.45$, $0.6$,and $0.75$). 
	Squares indicate the maximum shear rate in the nozzle channel under the flow conditions used in our simulations (cf. \sref{sec:methods-simulation-setups}). 
	(b) Corresponding velocity profiles for an average velocity of $\SI{5}{\milli\meter\per\second}$ inside the cylindrical printing nozzle.
	With increasing shear thinning strength, the velocity profile flattens at the center. 
	(c) Corresponding fluid stress profiles. 
	Stresses are linear with the radial position, and the maximum fluid stress decreases significantly with increasing $\alpha$ at constant flow rate.
%	\BF{ (because the pressure was adjusted to ensure a constant average flow rate regardless of alpha)}\SJM{see below eq (4)}
	}
	\centering
	% zplot-rheology-visc-velo-stress.plt
	\includegraphics[width=\linewidth]{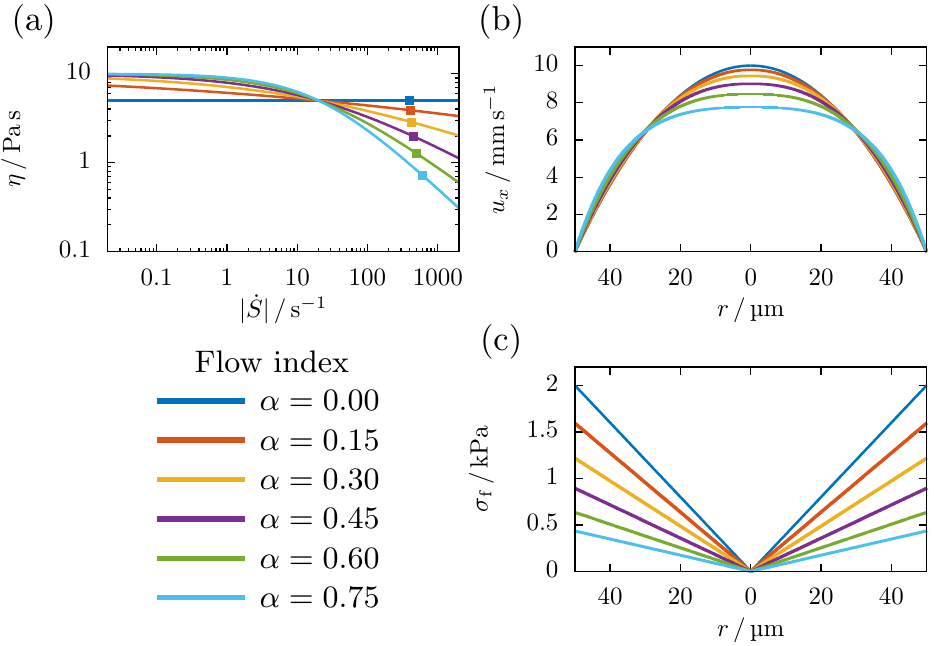}
\end{figure}

The shear thinning rheology is considered an essential material property for bioinks, as it serves two purposes: first, the large viscosity of the material at rest provides the necessary mechanical stability of the printed construct itself. Second, the shear thinning properties allow the material to be printed at significantly lower pressure --- considering the same printing speed ---, thus reducing the overall hydrodynamic stresses acting on cells inside the nozzle.
\\
%In our simulations in sections~\ref{sec:cell-cylinder}, \ref{sec:flow-analysis}, and \ref{sec:cell-nozzle}, we use six materials with increasing shear thinning behavior.
We describe the viscosity as function of the rate of strain $\rateofstrain$ according to a three-parameter simplified Carreau-Yasuda model, also known as Cross model \cite{muller_flow_2020-1,cross_rheology_1965}:
\begin{align}
\label{eq:viscosity-model}
\dynvisc\qty(\rateofstrain) = \frac{\dynvisc_0}{1 + \qty(K \rateofstrain)^\alpha }
\end{align}
Here, $\dynvisc_0$ is the zero-shear viscosity and the exponent $\alpha$ characterizes the shear thinning strength of the bioink, with $\alpha = 0$ for a Newtonian fluid, and $\alpha > 0$ for a shear thinning fluid.
The inverse of the time constant, $K\inverse$, defines the rate of strain at which the viscosity is equal to $\dynvisc_0/2$.
$\rateofstrain$ is calculated as the contraction of the rate of strain tensor $\dot{S}$ via
\begin{align}
\label{eq:rate-of-strain-contraction}
\rateofstrain = \sqrt{2\dot{S}_{ij}\dot{S}_{ij}}
\end{align}
with the tensor elements
\begin{align}
\dot{S}_{ij} = \frac{1}{2}\qty(\pdv{\velocity_i}{x_j} + \pdv{\velocity_j}{x_i}) \eqspace .
\end{align}
The diagonal elements of $\dot{S}$ are the rates of elongation of the fluid along the coordinate axes, and the off-diagonal elements are the respective shear rates.
%For the simple shear scenario as we use it in \sref{sec:methods-roscoe-linshear}, with $\dot{S}_{xy}$ being the only non-zero component of the rate of strain tensor, the rate of strain corresponds to the shear rate
%\begin{align}
%\dot{\gamma} = \rateofstrain = \pdv{u_x}{y} \eqspace .
%\end{align}
We choose $\dynvisc_0=\SI{10}{\pascal\second}$ and $K=\SI{0.05}{\second}$ for the zero-shear viscosity and the time constant, respectively.
This parameter choice roughly resembles the values obtained for $\SI{2.5}{\percent}$ alginate hydrogels \cite{hazur_improving_2020,muller_flow_2020-1,manojlovic_investigations_2006} which is a widely used bioink material.
In order to investigate the influence of the shear thinning strength in our calculations, we pick six different values for $\alpha$ between 0 and 1 with $ \alpha =0.75$~\cite{muller_flow_2020-1} corresponding to the said alginate solution.
The viscosity as function of the shear rate is depicted in \fref{fig-viscosity-shearrate-materials}(a).
For an idealized, i.e.\ infinitely long, cylindrical nozzle, the velocity profile and the fluid stress $ \sigma _ \mathrm{f}$ can be computed according to \cite{muller_flow_2020-1}\footnote{Or using our web tool under \url{https://bio.physik.fau.de/flow_webpage/flow.html} } as shown in \fref{fig-viscosity-shearrate-materials}(b and c) with the pressure adjusted so as to ensure the same flow rate for each $ \alpha $.
In our previous study \cite{muller_flow_2020-1}, we introduced this method to calculate the velocity, shear rate, viscosity, and shear stress, profiles for an inelastic shear thinning fluid in a cylindrical nozzle.
The central assumptions --- a laminar, uni-axial, pressure driven, flow --- are usually applicable for the description of bioink extrusion.
In the following, we define the fluid stress as:
\begin{align}
	\label{eq:fluidstress-definition}
	\fluidstress = \dynvisc\qty(\rateofstrain) \rateofstrain
\end{align}
We note that, if a constant extrusion pressure was used for calculation, the fluid stress profile in \fref{fig-viscosity-shearrate-materials}(c) would be the same regardless of $\alpha$~\cite{muller_flow_2020-1}.
\subsection{Cell elasticity}
\label{sec:methods-cell}
\subsubsection{Hyperelastic cell model}
\label{sec:methods-cell-hyperelastic-model}
Our cell is modeled as hyperelastic continuum, with a sphere as equilibrium configuration.
We provide extensive validation of the model in a previous publication \cite{muller_hyperelastic_2020-2}.
This includes AFM and FluidFM\textregistered\ measurements on biological cells and hydrogel particles as well as comparison to analytical theories \cite{roscoe_rheology_1967, gao_rheology_2011} and previous numerical simulations in shear flow \cite{rosti_rheology_2018,saadat_immersed-finite-element_2018}.
%This includes comparison with an axisymmetric model, and the deformation behavior of single cells \cite{gao_rheology_2011} as well as cell suspensions in linear shear flow \cite{rosti_rheology_2018,saadat_immersed-finite-element_2018}.
\\
As a hyperelastic model, we employ the neo-Hookean material model.
This model is strain-hardening for compressive strain, e.g. in AFM experiments, but also for shear strains as occurring mainly in microfluidic experiments. 
Its strain energy density  is computed via \cite[p. 100]{bower_applied_2010}
\begin{align}
\label{eq:neohookean-strainenergydensity}
U = \frac{\mu}{2} \qty(I-1) + \frac{\kappa}{2}\qty(J-1)^2 \eqspace ,
\end{align}
where $J=\det(F)$ is the determinant of the deformation gradient tensor $F_{ij}=\pdv{x_i}{y_j}$ \cite[p. 14, 18]{bower_applied_2010}, with the undeformed and deformed vertex coordinates $x_i$ and $y_i$, respectively.
$I= J^{-2/3}\tr(F\transpose F)$ denotes the second invariant of $F$.
As elastic parameters we choose a shear modulus of $\shearmodulus=\SI{1000}{\pascal}$ and a Poisson ratio of $\poissonratio=0.48$.
A simulation series with $\shearmodulus=\SI{500}{\pascal}$ is included in \sisref{9} of the Supplementary Material.
The Poisson ratio near $0.5$ provides sufficient incompressibility of the cell, while the shear modulus lies in the range typically found for mammalian stem cells \cite{kiss_elasticity_2011}.
In consistency with linear elasticity for small deformations, the shear and bulk modulus relate to the Young's modulus and Poisson ratio via
\begin{align}
\shearmodulus = \frac{\youngsmodulus}{2 \qty(1+\poissonratio)} \quad \mathrm{and} \quad \bulkmodulus=\frac{\youngsmodulus}{3 \qty(1-2\poissonratio)} \eqspace .
\end{align}
The cell radius is chosen as $\cellradius=\SI{8}{\micro\meter}$ ($6$ in simulation units), and the mesh consists of $9376$ tetrahedra.

In our numerical method, the interior of the cell is filled with the same fluid as the outside fluid.
Together with the Neo-Hookean elasticity, this leads to an effectively viscoelastic cell response \cite{muller_hyperelastic_2020-2}.
\subsubsection{Force calculation and flow coupling}
\label{sec:methods-cell-force}
For numerical simulations, the spherical volume is uniformly tetrahedralized using the meshing software Gmsh \cite{geuzaine_gmsh_2009}.
The elastic forces acting on each vertex of one tetrahedron are obtained via differentiation of the strain energy density \eqref{eq:neohookean-strainenergydensity} with respect to the relative vertex displacement,
\begin{align}
	f_i = -V_0 \pdv{U}{u_i} \eqspace ,
\end{align}
where $V_0$ denotes the reference volume of the tetrahedron and $u_i=y_i-x_i$.
This approach is explained in detail in section~$3.1$ in \cite{muller_hyperelastic_2020-2}.
\\
The coupling between the cell model and the bioink is realized using an immersed-boundary algorithm \cite{mittal_immersed_2005,peskin_immersed_2002}.
After computation of the cell vertex forces, they are transmitted into the fluid via extrapolation from the moving Lagrangian cell mesh onto the static Eulerian Lattice Boltzmann grid.
The two-way coupling is completed through advecting the cell vertices with the local interpolated fluid velocity.
%With this algorithm applied to all cell vertices, particularly including those inside the cell volume, it becomes effectively viscoelastic \cite{muller_hyperelastic_2020-2}, with an internal viscosity equal to that of the surrounding fluid.
%
\subsubsection{Cell stress calculations}
\label{sec:methods-cell-stress}
In addition to the elastic forces, we are able to obtain the internal stress distribution inside our cell model.
We compute the Cauchy stress tensor in each tetrahedron from the strain energy density and the deformation gradient tensor according to Bower \cite[p. 97]{bower_applied_2010} as:
\begin{align}
\label{eq:cauchy-stress-general}
\stress_{ij} = J^{-1} F_{ik} \pdv{U}{F_{jk}}
\end{align}
For the neo-Hookean model in \eqref{eq:neohookean-strainenergydensity}, this becomes
\begin{align}
\label{eq:cauchy-stress-nh}
\stress_{ij} = \frac{\shearmodulus}{J^{5/3}}\qty( B_{ij} - \frac{1}{3} B_{kk} \kronecker{ij} ) + \bulkmodulus \qty(J-1) \kronecker{ij} \eqspace ,
\end{align}
where $B = F F\transpose$ denotes the left Cauchy-Green deformation tensor.
\\
In order to obtain a simple scalar observable to quantify the cell stress, we start from the local von Mises stress in each tetrahedron.
The von Mises stress is an invariant of the Cauchy stress tensor and commonly used in plasticity theory to predict yielding of materials under multiaxial loading conditions through construction of a fictitious uniaxial loading.
Using the principal stresses, \ie, the eigenvalues $\stress_1,$ $\stress_2$, and $\stress_3$, of the Cauchy stress tensor \eqref{eq:cauchy-stress-nh}, we calculate \cite[p. 48]{bower_applied_2010}
%\SG{ can we remove the superscript $ \mathrm{tet}$ here? It would fit better with the new section on cell stress }
\begin{align}
\label{eq:vonMises-stress}
\vonmises  = \sqrt{\frac{1}{2}\qty[ \qty(\stress_1 - \stress_2)^2 + \qty(\stress_2 - \stress_3)^2 + \qty(\stress_3 - \stress_1)^2 ]}.
\end{align}
An alternative equivalent formulation to \eqref{eq:vonMises-stress} is the contraction of the deviator of the Cauchy stress tensor $\stress_{ij}^\mathrm{dev} = \stress_{ij} - \frac{1}{3}\stress_{kk}$. It reads:
\begin{align}
\label{eq:vonMisesContract}
\vonmises  = \sqrt{\frac{3}{2}\stress_{ij}^\mathrm{dev}\stress_{ij}^\mathrm{dev}}.
\end{align}
The total cell stress $\vonmises$ is then computed by averaging the local von Mises stress over all tetrahedra in the cell model weighted by the undeformed volume of each tetrahedron.

%Two states of the same distortion energy have an equal von Mises stress.\TODO{from wiki}
%We define the cell stress $\vonmises$ \SG{ symbol  }as the von Mises stress averaged over all tetrahedra in the cell model weighted by reference tetrahedron volume. \SG{ true? }
%\\
%\SG{ sollte dieser Abschnitt nicht woanders hin? Sind das methods? }\TODO{TODO!}\SJM{$\rightarrow$}From \eqref{eq:cauchy-stress-nh}, the Cauchy stress can be calculated for a triaxial elongation of an incompressible cell ($J=1$).
%The deformation gradient tensor of a sphere deformed into an ellipsoid is $F_{ij} = \alpha_i \kronecker{ij}$, with $\alpha_i$ denoting the cell strains along the coordinate axes.
%The non-zero diagonal elements of the Cauchy stress are then found as
%\begin{align}
%\label{eq:cell-stress-triaxial}
%\stress_{11} & = \frac{\shearmodulus}{3} \qty(2\alpha_1^2 - \alpha_2^2 - \alpha_3^2) \eqspace ,
%\end{align}
%with similar expressions for $\stress_{22}$ and $\stress_{33}$ obtained by cyclic change of indices.\SJM{$\leftarrow$}
%
%\subsubsection{Validation of the cell stress calculation}
\subsubsection{Validation of the cell stress calculation}
\label{sec:methods-roscoe-linshear}
%An analytical theory describing the deformation and stresses of a cell embedded in linear flow was proposed by Roscoe \cite{roscoe_rheology_1967}, based on the work of Jeffery \cite{jeffery_motion_1922}.
%\\
We validate our cell stress calculations using a linear shear flow setup:
the simulation box with dimensions $10 \times 15 \times 5$ ($x \times y \times z$ in units of $\cellradius$) is bounded by two planes at $y=0$ and $y=15\cellradius$, moving with a tangential velocity in $\pm x$-direction.
This creates a linearly increasing velocity across the gap and thus a constant shear rate $\dot{\gamma}$ in the box.
%The shear rate is varied to achieve a range of Capillary numbers up to $1.5$, while the fluid viscosity ($\alpha=0$) and the cell's shear modulus remain constant.
The shear rate is varied to achieve a range of fluid stresses up to $\SI{1.5}{\kilo\pascal}$, while the fluid viscosity ($\alpha=0$) and the cell's shear modulus remain constant.
In non-dimensional terms, this range corresponds to capillary numbers $\capillarynumber=\frac{\fluidstress}{\shearmodulus}$ between $\num{0}$ and $\num{1.5}$.

\begin{figure}
	\centering
	\caption{\label{fig-roscoe-shear-validation}Comparison of the cell stress predicted by Roscoe \cite{roscoe_rheology_1967} and the average cell stress of our model in shear flow. 
	Insets show the stationary, tank-treading shape of the simulated cell at fluid stresses corresponding to $\capillarynumber=0.2$, $0.6$, and $1.2$.
}
	\includegraphics[width=\linewidth]{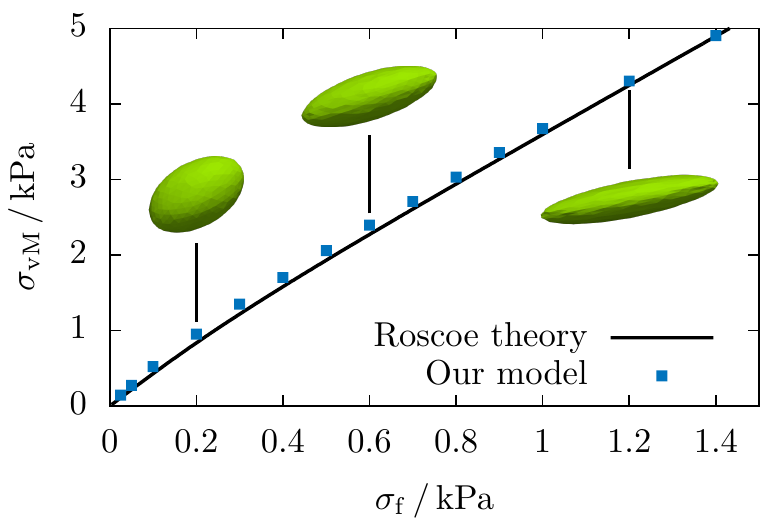}
\end{figure}

During the simulation, the initially spherical cell traverses through a series of ellipsoidal deformations before reaching a stationary state, at which the whole cell volume performs a tank-treading motion, \ie, the cell vertices rotate around the fixed ellipsoidal cell shape.
In \fref{fig-roscoe-shear-validation}, we compare the elastic cell stress in the stationary state calculated by \eqref{eq:vonMises-stress} to the analytical calculations of Roscoe \cite{roscoe_rheology_1967} (detailed in \sisref{4}) and find excellent agreement for a realistic range of fluid stresses.

%\SG{ check reformulation } \SJM{$\checkmark$}
In addition to the elastic stress, we compute the viscous contribution resulting from the fluid motion enclosed by the cell volume.
This quantity is extracted from the Lattice-Boltzmann strain rate tensor field~\cite{chai_multiple-relaxation-time_2011,muller_flow_2020-1} inside the cell using our method from~\cite{lehmann_efficient_2020} and averaging over the cell volume.
In \sifref{2} we show that the agreement of the numerically obtained viscous cell stress with Roscoe theory is equally good as for the elastic component.
\\
We note that cell and fluid stress in \fref{fig-roscoe-shear-validation} are time-independent and stationary.
We further dissect their relation in detail in \sref{sec-dissection-cell-stress}.

\subsection{Bioprinting simulations}
\label{sec:methods-simulation-setups}
The two essential parts of the bioprinting process are (i)~the flow inside the nozzle channel and (ii)~the flow transition at the nozzle exit.
Both situations will be studied separately in this work.

\subsubsection{Nozzle channel}
\label{sec-nozzle-channel-setup}
We model the nozzle channel using a periodic cylindrical no-slip channel with a radius of $R=\SI{50}{\micro\meter}$ and length of $\SI{133}{\micro\meter}$ ($37.5$ and $100$ in simulation units), as depicted in the left dashed box in \fref{fig-setup-schematics}. 
The shear thinning fluid dynamics are solved by the Lattice-Boltzmann method as described in section~\ref{sec:methods-fluid}.
No-slip boundary conditions are imposed at the channel wall.
The flow is driven by a pressure gradient $\pgrad$ along the nozzle axis.
To compare the different bioinks detailed in \sref{sec:methods-bioink-rheology}, we consider a fixed average velocity of $\SI{5}{\milli\meter\per\second}$ (volumetric flow rate $ \flowrate= \SI{3.93e-11}{\cubic\meter\per\second}\approx\SI{141}{\micro\litre\per\hour}$).
The corresponding pressure gradient is different for each $ \alpha $ and is obtained according to \cite{muller_flow_2020-1}.
Our input parameters as well as averaged and maximum quantities of the nozzle channel flow are listed in \tref{tab-nozzle-channel-parameters}.
We note that compared to common flow cytometry setups \cite{fregin_high-throughput_2019,otto_real-time_2015}, the channel radius in typical bioprinting applications is significantly larger, thus allowing cells to flow off-centered.
To account for this, a single spherical cell is inserted at different radial starting positions of $\num{0}$, $\num{1.5}\,\cellradius$, $\num{3}\,\cellradius$, and $\num{4.5}\,\cellradius$, as shown in \fref{fig-setup-schematics}.

\subsubsection{Nozzle exit}
\label{sec-nozzle-exit-setup}
\begin{figure}[h]
	\centering
	\caption{\label{fig-setup-schematics}Schematic of the bioprinter setup: the nozzle channel is bounded by a cylindrical wall and periodic in $x$-direction. Single cells are inserted at different radial offsets. The nozzle exit consists of the transition from the no-slip nozzle channel to the free bioink strand. Cells are initialized in a deformed state close to the transition. }
	\includegraphics[width=\linewidth]{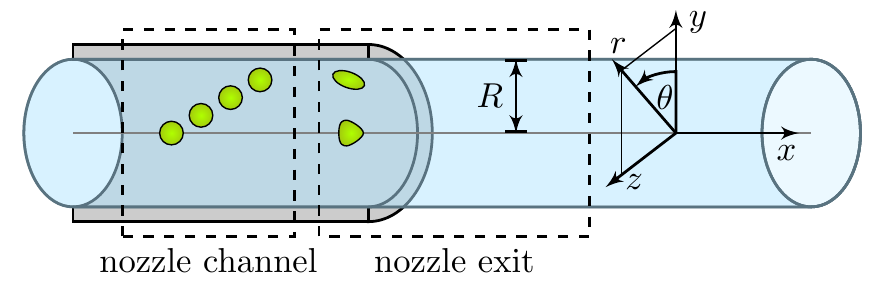}
\end{figure}

The geometry of our simulations at the nozzle exit is depicted by the right dashed box in \fref{fig-setup-schematics}, the flow dynamics are again solved by the Lattice-Boltzmann method. 
%The free bioink strand behind the nozzle exit is assumed to have the same radius, with free-slip boundary conditions applied at the fluid surface which result in a plug motion of the bioink.
The free bioink strand of length $\SI{933}{\micro\meter}$ ($700$ in simulation units) behind the nozzle exit is assumed to have the same radius as the inner radius of the nozzle channel, with free-slip boundary conditions applied at the fluid surface, which result in a plug motion of the bioink.
This way we neglect the small extension of the bioink strand at the nozzle exit known as Barus effect or die swell~\cite{fisch_improved_2021}.
Equal flow conditions as inside the nozzle channel are achieved by applying the average velocity of $\SI{5}{\milli\meter\per\second}$ as normal velocity at the circular inflow and outflow planes, instead of a constant pressure gradient as used in the nozzle channel setup.
We insert a single cell at different radial positions as explained above. 
The starting configuration of the cell is taken from the corresponding simulation of the nozzle channel setup, \ie, the cell is inserted in a deformed state close to the nozzle exit, as shown in the first frames in \fref{fig-nozzle-cellstress-newtonian}.
\begin{table}
	\caption{\label{tab-nozzle-channel-parameters}Flow parameters of our bioprinter setup with a nozzle radius of $\SI{50}{\micro\meter}$ and an average velocity of $\SI{5}{\milli\meter\per\second}$.}
	\begin{tabular}{r r r r r r r}
%		$\alpha$ & $\pgrad\,/\,\si{\pascal\per\meter}$ & $\fluidstress^\mathrm{max}\,/\,\si{\kilo\pascal}$ & $\fluidstress^\mathrm{avg}\,/\,\si{\kilo\pascal}$ & $\velocity_x^\mathrm{max}\,/\,\si{\milli\meter\per\second}$ & $\shearrate^\mathrm{max}\,/\,\SI{e2}{\per\second}$ & $\shearrate^\mathrm{avg}\,/\,\SI{e2}{\per\second}$ \\ \hline \\
		$\alpha$ & $\pgrad$ & $\fluidstress^\mathrm{max}$ & $\fluidstress^\mathrm{avg}$ & $\velocity_x^\mathrm{max}$ & $\rateofstrain^\mathrm{max}$ & $\rateofstrain^\mathrm{avg}$ \\
		$-$ & $\si{\pascal\per\meter}$ & $\si{\kilo\pascal}$ & $\si{\kilo\pascal}$ & $\si{\milli\meter\per\second}$ & $\si{\per\second}$ & $\si{\per\second}$ \\ \hline &&&&&&\\
		$\num{0.00}$ & $\num{8.00e7}$ & $\num{2.00}$ & $\num{1.33}$ & $\num{10.0}$ & $\num{400}$ & $\num{267}$ \\
		$\num{0.15}$ & $\num{6.37e7}$ & $\num{1.59}$ & $\num{1.06}$ & $\num{9.77}$ & $\num{410}$ & $\num{265}$ \\
		$\num{0.30}$ & $\num{4.87e7}$ & $\num{1.22}$ & $\num{0.81}$ & $\num{9.45}$ & $\num{426}$ & $\num{262}$ \\
		$\num{0.45}$ & $\num{3.58e7}$ & $\num{0.89}$ & $\num{0.60}$ & $\num{9.03}$ & $\num{454}$ & $\num{257}$ \\
		$\num{0.60}$ & $\num{2.54e7}$ & $\num{0.63}$ & $\num{0.42}$ & $\num{8.47}$ & $\num{502}$ & $\num{251}$ \\
		$\num{0.75}$ & $\num{1.74e7}$ & $\num{0.44}$ & $\num{0.29}$ & $\num{7.77}$ & $\num{604}$ & $\num{242}$ \\
	\end{tabular}
\end{table}
\section{Results}
\label{sec:results}
\subsection{Dissecting the notion of ''cell stress''}
\label{sec-dissection-cell-stress}
In many situations, it has become a common approach to invoke the term ``cell stress'' and to equate it directly to the fluid stress, \ie, the viscosity multiplied by the local shear rate at the cell position.
Here, we show that this simple approach, while being correct in its order of magnitude, hides a good amount of the more complex features of intracellular stress.
To illustrate this, we apply the theory of Roscoe \cite{roscoe_rheology_1967} for a cell in linear shear flow, which accurately describes cell behavior in our numerical simulations (see section \ref{sec:methods-roscoe-linshear}) and in microchannel experiments~\cite{gerum_viscoelastic_2022}, provided that the cell does not flow in the channel center where the shear rate approaches zero.

Inside a flowing cell, two qualitatively different kinds of stress arise.
The first kind are viscous stresses that are caused by frictional motion (\ie\  tank-treading) of the cell interior.
The second kind are elastic stresses that are caused by the deformation (\eg\  shearing and stretching) of the cell.
The magnitude of the former are governed by the cell's internal viscosity, while the latter are set by its elastic moduli.
We note that, in principle, both a cell's viscosity and its elasticity can be non-homogeneous, \ie, they vary spatially throughout the cell, and anisotropic, \ie, they depend on direction, \eg, due to alignment of certain cytoskeletal elements.
Here and in most other situations, these more complicated effects are neglected, and the cell is considered a homogeneous, isotropic viscoelastic medium.
Furthermore, as shown in \cite{roscoe_rheology_1967} for a cell in pure shear flow, stability requires that viscous and elastic cellular stresses do not vary between different locations inside the cell.
Their value can be calculated from Roscoe theory as detailed in the Supporting Information (eqs.~\sieqref{43} and \sieqref{48}).

\begin{figure}
	\centering
	\caption{\label{fig-limits-cellstress}Components of the cell stress tensor $\stress_{ij}$ normalized by the undisturbed fluid stress $\fluidstress$ across multiple orders of magnitude of the Capillary number computed using the Roscoe theory.
	Elastic stresses are shown in blue, viscous stresses are shown in purple. 
	Components not appearing in the figure remain zero throughout.
	The directions $1,2,3$ refer to a body-fixed coordinate system as indicated by the insets.}
	\includegraphics[width=\linewidth]{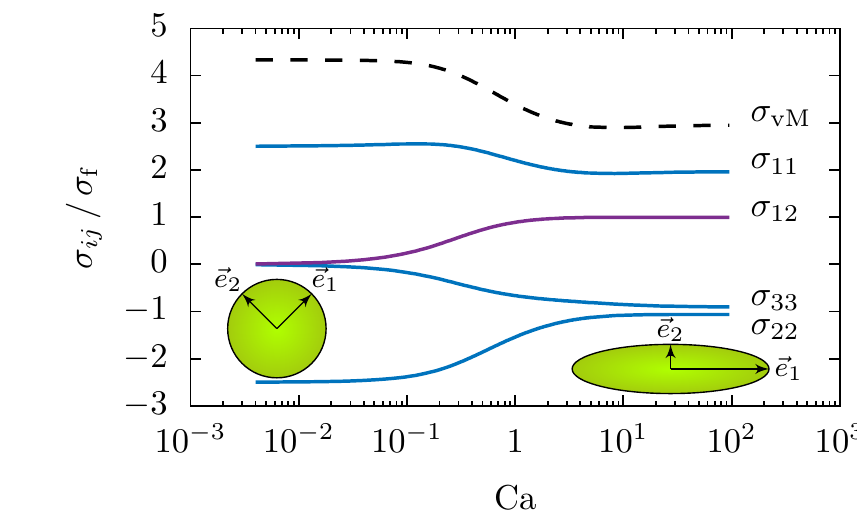}
\end{figure}

We start with the limiting case of low shear rates corresponding to small capillary numbers $ \mathrm{Ca} = \frac{ \sigma _f } { \mu  } \to 0$.
In this limit, fluid stresses are not sufficient to cause significant cell deformation, and the cell essentially remains spherical.
Indeed, the classical calculation for a rigid sphere in shear flow detailed in \sisref{6} of the Supporting Information yields a surprisingly accurate description of this limit.
The cell rotates as a rigid body, which implies the absence of internal frictional motions and thus leads to a vanishing viscous cell stress as shown by the purple curve in \fref{fig-limits-cellstress}.
Similarly, elastic stresses in the vorticity direction vanish as shown by the $ \sigma_{33}$ component in \fref{fig-limits-cellstress}.
A positive stress appears in a direction inclined by $\SI{45}{\degree}$ with respect to the flow direction ($ \sigma _{ 11 }$), with a corresponding negative stress in the perpendicular direction.
Their magnitude is precisely $5/2$ times the undisturbed fluid stress $ \sigma _ \mathrm{f}$, which exactly corresponds to the situation of the rigid sphere as shown in \sisref{6}. 

In the opposite limit of high shear rates ($ \mathrm{Ca}\to\infty$), the situation becomes more involved.
In agreement with our numerical simulations shown in \fref{fig-roscoe-shear-validation}, the cell becomes strongly elongated and aligned in flow direction.
Due to the persisting tank-treading motion, internal viscous stresses do not disappear.
Instead, the flatness of the cell shape minimizes the cell's disturbing influence on the surrounding fluid flow, and indeed the cell's internal viscous stress now becomes equal to the undisturbed fluid stress, as can be seen by the purple curve in \fref{fig-limits-cellstress}.
Maintaining the flattened cell shape, however, in addition requires elastic stresses.
As shown by the blue curve in \fref{fig-limits-cellstress}, all three elastic stress components arise with their ratios being $ \sigma _{ 11 }: \sigma _{ 22 } : \sigma _{ 33 }=2:-1:-1$.
The positive stress in flow direction, $ \sigma _{ 11 }$ is balanced by negative stresses in the two other directions.
These ratios can easily be understood by the analogy with a uniaxially stretched beam as detailed in \sisref{7} of the Supporting Information.

Despite this complexity, it may be helpful in many situations to have at hand a single measure to quantify ``cell stress''.
Such a measure can be provided by the elastic von Mises stress $\vonmises$ given in \eqref{eq:vonMisesContract} which we include as the black dashed line in \fref{fig-limits-cellstress}.
The ratio $ \vonmises / \sigma _ \mathrm{f}$ transitions from $\frac{5}{2}\sqrt{3}$ at low to $3$ at high $\capillarynumber$.
In the intermediate range, the proportionality factor is situated between these two limits.
As can also be deduced from \fref{fig-roscoe-shear-validation}, the relation between $\vonmises$ and $\fluidstress$ changes the most in the range of $0.1 < \capillarynumber < 1$, while otherwise an approximately linear dependency emerges.

From the results of this subsection, we conclude that the common approach of equating (undisturbed) fluid stress to ``cell stress'' can be a reasonable approximation for low and high Capillary numbers.
%Observing that the ratio $ \vonmises / \sigma _ \mathrm{f}  \approx 1$ for all capillary numbers demonstrates that the common approach of equating (undisturbed) fluid stress to ''cell stress'' is a reasonable, but not an exact, approximation.
%
\subsection{Cell flowing inside the nozzle channel}
\label{sec:cell-cylinder}
\begin{figure}
	\centering
	\caption{\label{fig-cylinder-cell-center-stress}Internal stress distribution for a cell flowing at the center of the nozzle channel in (a) a Newtonian fluid and (b) a strongly shear thinning fluid. 
	(c) Internal stress distribution and radial migration of an off-centered cell towards the axis of the nozzle channel in a Newtonian bioink. 
	The bottom labels give the axial distance traveled during the time given by the top labels.
	}
	\includegraphics[width=\linewidth]{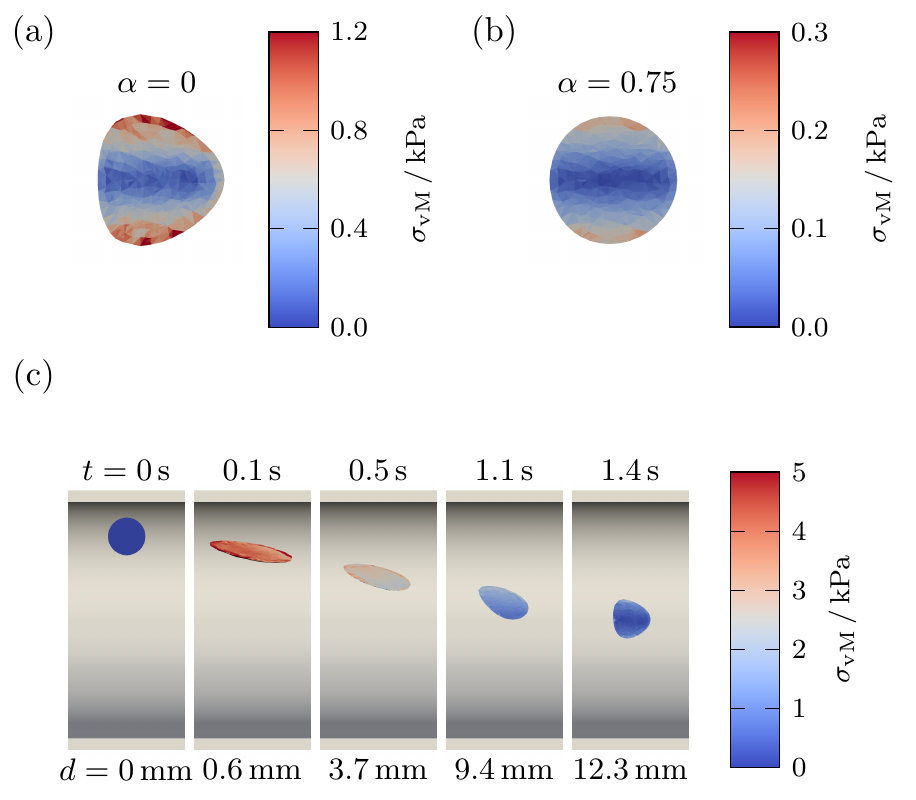}
\end{figure}

\begin{figure}
	\centering
	\caption{\label{fig-cylinder-time-posy-materials} 
	Radial migration $r$ (center-of-mass) of a cell starting near the nozzle wall for different shear thinning strengths (see (d) for color labels) with (a) constant flow rate and (b) constant pressure gradient ($\pgrad=\SI{1.14e+07}{\pascal\per\meter}$). 
	For constant $G$,  the migration speed is almost independent of $\alpha$. 
	(c,d)~Cell stress as function of the local fluid stress compared to the Roscoe theory (black line). Due to the radial migration of the cell, the cell experiences a continuous change of the local fluid stress over time. (c)~The cell starting at different offsets (from right to left: $4.5\cellradius$, $3\cellradius$, and $1.5\cellradius$) in the Newtonian fluid. The duration of the deformation from the spherical reference shape to the approximately elliptical shape is given by the points. (d)~Cells starting at offset $4.5\cellradius$ for bioinks with increasing shear thinning strength in comparison with the Roscoe theory. For details, see text.
%	\SG{ The figre in the SI has the nice $r/R_c$ axis on top. Please include here as well. }	\SJM{$\checkmark$}
	%\SJM{The $r/R_c$ axis makes sense only for subfigure c. subfigure d has six different axes for $r/R_c$.}
}
	%($\pgrad=\SI{1.141535e+07}{\pascal\per\meter}$)
	% zplot-cylinder-time-radialpos.plt
	\includegraphics[width=\linewidth]{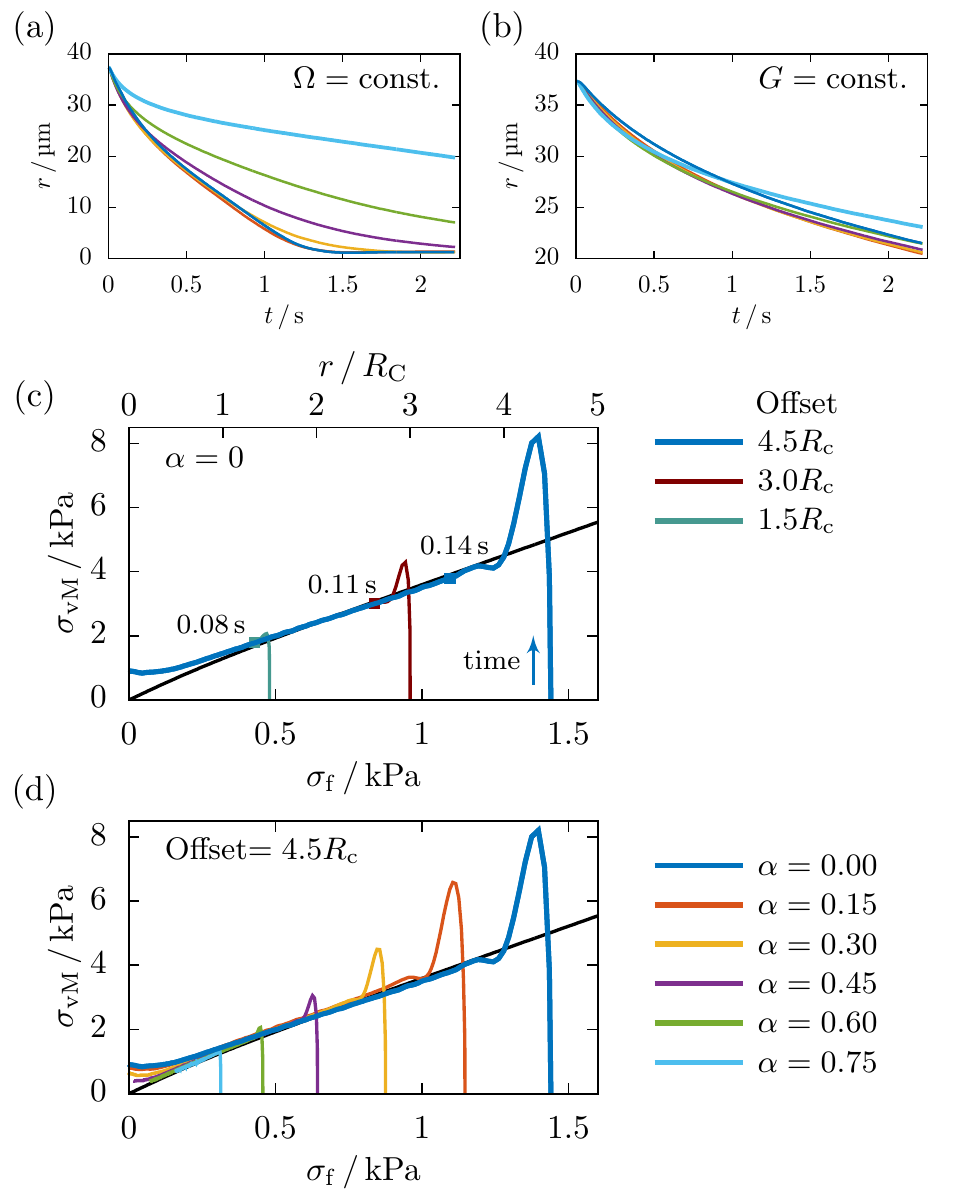}
\end{figure}

Using our setup described in \sref{sec-nozzle-channel-setup}, we investigate the cell behavior and the cell's internal stress distribution during the flow inside the nozzle.
Depending on the initial radial position, we observe two modes of deformation of the cell:
\\
(i) A cell flowing along the axis of the nozzle channel assumes an axisymmetric bullet-like shape, as can be seen in \fref{fig-cylinder-cell-center-stress}(a) and (b) for a Newtonian and a highly shear thinning bioink, respectively.
In both cases, the radial dependency of the internal cell stress is highly non-homogeneous and resembles the linearly increasing fluid stress of the undisturbed liquid (cf. \fref{fig-viscosity-shearrate-materials}(c)), since the cell's surface has to balance higher fluid shear stresses for increasing $\radial$ in the stationary state.
%\SG{ This sounds good, but why does the same not happen when the cell is off-center? } \SJM{It does. That is why the cell shape is not perfectly ellipsoidal.}
However, the magnitude of the stress and likewise the cell deformation decrease significantly when the shear thinning index $ \alpha $ is increased at the same volumetric flow rate.
This finding may explain earlier experimental observations in which more shear thinning bioinks were found to increase cell survival in bioprinting \cite{ouyang_effect_2016,billiet_3d_2014} when the pressure was reduced to ensure equal flow rates for all conditions.
\\
(ii) A cell flowing off-center deforms into an approximately ellipsoidal shape exhibiting tank-treading motion.
Due to the curvature of the flow, the cell migrates towards the channel center (sometimes referred to as margination), where it eventually assumes the bullet-like shape discussed in the previous paragraph.
A sequence of simulation snapshots for a cell flowing in the Newtonian bioink is shown in \fref{fig-cylinder-cell-center-stress}(c), where the internal stress distribution of the off-centered cells can be observed.
\Fref{fig-cylinder-time-posy-materials}(a) shows the corresponding development of the radial position over time starting from an offset of $\SI{36}{\micro\meter}$.
With increasing shear thinning strength, \ie, decreasing pressure gradient, the cell takes longer to migrate towards the channel center.
\\
In \fref{fig-cylinder-time-posy-materials}(b) the same situation is studied for a constant pressure gradient.
We find that here the migration speed of the cell becomes independent of the shear thinning properties of the bioink and thus conclude that cell migration is determined predominantly by the applied pressure gradient and not the flow speed.
This finding can be understood since the stress, and thus the forces, that the cell feels are directly determined by the local fluid stress.
%This finding can be understood as follows:
%Considering first a cell-free system, applying the same pressure gradient creates the exact same radial shear stress profile inside the nozzle, independent of the bioink rheology~\cite{muller_flow_2020-1}.
%Since a cell is small compared to the nozzle radius, its presence does not significantly alter this shear stress profile.
%Hence the cell experiences similar forces exerted on its surface which lead to similar deformations and, ultimately, similar elastic restoring forces are transmitted back into the fluid.
%Since this coupling between elastic forces of the cell and viscous forces of the bioink determines the cell motion~\cite{chen_inertia-_2014}, neither the rheology of the bioink nor the magnitude of the flow velocity determine the radial motion of the cell.
Therefore, when printing bioinks with different rheology at the same printing pressure, the radial cell distribution will not change.
When printing bioinks with increasing shear thinning strength at the same flow rate, by contrast, fewer cells will migrate to the center of the nozzle.
\\
The ellipsoidal cell shape during the migration allows us to compare the cell stress to the prediction of the Roscoe theory \cite{roscoe_rheology_1967} detailed in \sref{sec:methods-roscoe-linshear}.
In \fref{fig-cylinder-time-posy-materials}(c) we plot the development of the cell stress in a Newtonian bioink when cells are initially placed at different offsets from the nozzle center.
Due to the migration of the cells towards the channel center, the local fluid stress experienced by a given cell decreases monotonically over time.
In order to directly compare with the prediction of Roscoe theory, which assumes a constant fluid shear stress, we choose this local fluid stress as abscissa.
%Cells start at offsets of $1.5 R_ \mathrm{c}$, $3.0 R _{ \mathrm{c} }$ and $4.5 R_ \mathrm{c}$ corresponding to initial fluid stresses of $ \sigma _ \mathrm{f} \approx \SI{0.5}{\kilo\pascal}$, $\SI{1.0}{\kilo\pascal}$, and $\SI{1.5}{\kilo\pascal}$, respectively.
Cells start at offsets of $\SI{12}{\micro\meter}$, $\SI{24}{\micro\meter}$ and $\SI{36}{\micro\meter}$ corresponding to initial fluid stresses of $ \sigma _ \mathrm{f} \approx \SI{0.5}{\kilo\pascal}$, $\SI{1.0}{\kilo\pascal}$, and $\SI{1.5}{\kilo\pascal}$, respectively.
The initial shape is undeformed and thus $ \vonmises =0$ for $t=0$.
The cell first experiences a transient of large stresses and quickly relaxes towards the cell stress predicted by Roscoe where the curved flow is locally approximated as a pure shear flow, as indicated by the square symbols.
Due to the migration towards the channel center, the cell stress decreases with time and radial position.
The curves of all initial radial offsets perfectly agree with the prediction of the Roscoe theory, as long as the cell's radial position is larger than $\cellradius$.
When the cell is close to the channel center, the local shear flow approximation becomes invalid, thus causing deviations from the theoretical prediction.
%Considering now the shear thinning properties of the bioink, we plot in \fref{fig-cylinder-fluidstress-cellstress-materials}(b) the cell stress of cells starting at offset $4.5\cellradius$ for different $\alpha$ in comparison with the Roscoe theory.
\\
A similar plot is provided for shear thinning bioinks in \fref{fig-cylinder-time-posy-materials}(d) where the stress of cells starting at offset $4.5\cellradius$ for different $\alpha$ is compared with Roscoe theory.
We again find excellent agreement with the Roscoe theory independent of the shear thinning strength.
This finding may seem surprising at first, as the theory of Roscoe is designed for purely Newtonian fluids surrounding the cell, but stays valid for shear thinning bioinks as well.
This demonstrates that the key property determining cell motion is indeed not the shear rate, but rather the shear stress.
The plots for the remaining cell offsets and bioinks are included in the SI (cf.~\sifref{3}).
\subsection{Analysis of the flow field at the nozzle exit}
\label{sec:flow-analysis}
\begin{figure}
	\caption{\label{fig-nozzle-exit-profiles}$x$-$y$-slices of the flow at the nozzle exit for increasingly shear thinning fluids: 
	(a)~axial velocity component and 
	(b)~radial velocity component with streamlines as overlay. 
	(c)~The ratio of maximum axial velocity inside the nozzle to the average flow velocity as function of the shear thinning strength $\alpha$. 
	(d)~The ratio of maximum radial velocity after the nozzle exit to the average flow velocity as function of $\alpha$.
%	\SG{ what are the black squares? } \SJM{simulation series with just fluid for more $\alpha$ values, not done for cells}\SG{ then I would make them all black. In fact, the colors are not explained here anyway }
%	\BF{I would have predicted that the strand extends a bit when it exits the nozzle. This is also what we see when wi image it. } \SJM{That is a consequence of our simulation method, as we constrain the bioink strand thickness. While it is true that we would expect some strand widening, it is hard to incorporate in a systematic manner, because we cannot quantify the strand widening as function of the shear thinning strength (or even elastic bioink properties). This restriction therefore describes the ``ideal'' case of perfect shape fidelity with controlled strand thickness.}
	}
	\centering
	\includegraphics[width=\linewidth]{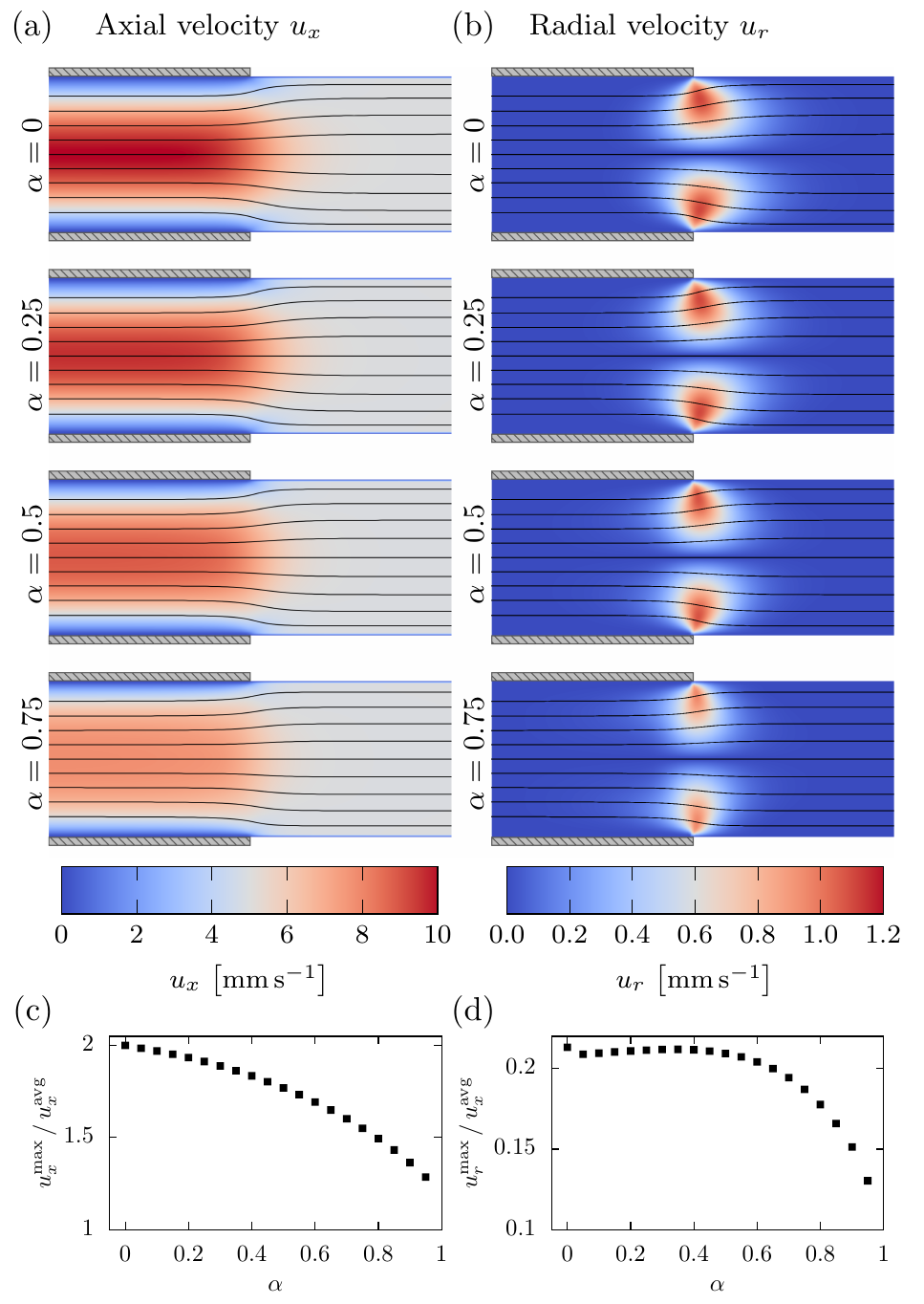}
\end{figure}

\begin{figure}
	\caption{\label{fig-nozzle-flow-deriv-decomp-newtonian}Decomposition of the fluid stress in shear \eqref{eq-nozzle-fluidstress-decomposition-shear} and elongational component \eqref{eq-nozzle-fluidstress-decomposition-elong} at the nozzle exit for the (a-d) Newtonian fluid and (e-h) the highly shear thinning bioink with $\alpha=0.75$. 
	%$d_x$ denotes the axial ($x$-) distance from the nozzle orifice.
	$x$ denotes the axial distance from the nozzle orifice. 
%	\SG{ Ich würde $ \sigma _ \mathrm{f}$ links und rechts die gleiche Farbe geben. Sonst sieht es so aus, als wären andere Größen. Dabei ist ja nur das eine shear-thinning und das andere nicht, oder? } \SJM{ich würde gern die Farbkodierung beibehalten, da die in allen Plots mit $\alpha$ benutzt wird.} \SG{ ok }
		%From top to bottom, the data is given for the radial position $0$, $1.5\cellradius$, $3\cellradius$, and $4.5\cellradius$.
	}
	\centering
	% zplot-nozzle-flow-stress-decomposition.plt
	\includegraphics[width=\linewidth]{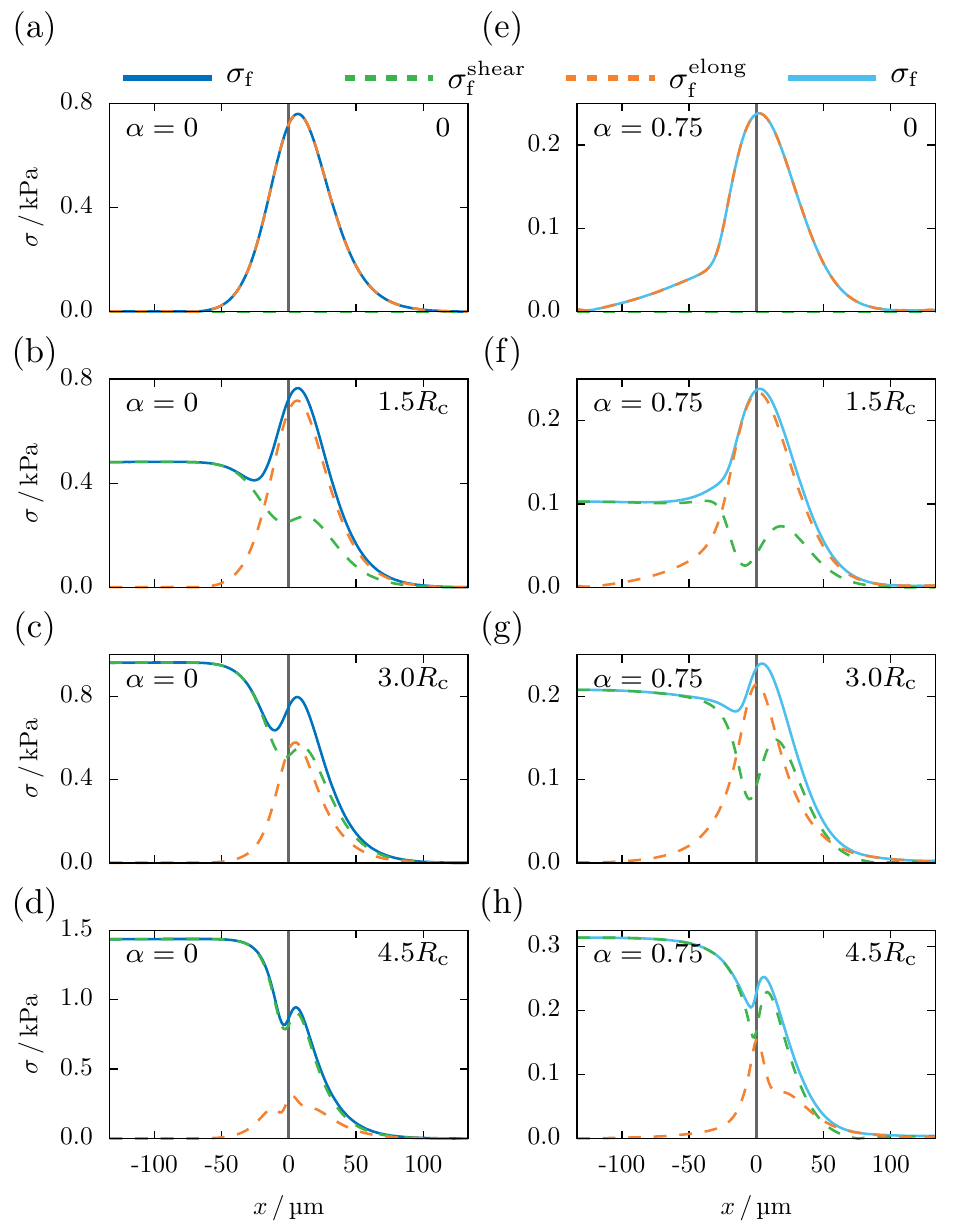}
\end{figure}

In this section, we investigate the influence of the shear thinning rheology of the bioinks introduced in \sref{sec:methods-bioink-rheology} on the undisturbed (cell free) flow field at the nozzle exit, where the transition from nozzle channel to the free bioink strand causes additional radial flows.
We use the second setup described in \sref{sec:methods-simulation-setups}, without a cell, and run the calculations until the flow becomes stationary.
%
% When passing the nozzle exit, additional radial flow components arise from the transition from the bounded channel to the free bioink strand. These radial flows as well as the concurrent change in axial velocity affect the cell for a much shorter time span compared to the flow inside the nozzle channel. While flowing inside the nozzle channel, the cell is subjected to the hydrodynamic forces for the longest time span, \ie, the residence time \cite{paxton_proposal_2017}, which depends on the (local) flow velocity and the length of the nozzle.
\\
In \fref{fig-nozzle-exit-profiles}(a) and (b), we show $x$-$y$-slices of the velocity profiles for the axial and radial velocity, respectively, at different values of the shear thinning parameter $ \alpha $.
From top to bottom, the shear thinning strength of the fluid increases, while the flow rate is kept constant.
The axial velocity component in \fref{fig-nozzle-exit-profiles}(a) shows the same trend for increasing $\alpha$ as seen in \fref{fig-viscosity-shearrate-materials}(b):
the flow develops a central plateau inside the nozzle channel which at the nozzle exit transitions into the plug flow inside the bioink strand.
Indeed, as shown in \fref{fig-nozzle-exit-profiles}(c), the ratio $\velocity_x^{\mathrm{max}}/\velocity_x^{\mathrm{avg}}$ between the maximum velocity inside the nozzle channel and the average velocity assumes the Poiseuille value of 2 at $ \alpha =0$ and decreases towards the plug-flow value of 1 for increasing shear thinning strength.
\\
The second column, \fref{fig-nozzle-exit-profiles}(b), shows the corresponding radial flow components. 
Due to the radial symmetry, they vanish at the center and increase towards the boundary, showing a drop-like shape with its tip pointing to the position of the nozzle orifice, where the boundary conditions change.
The radial flow components decrease with increasing $\alpha$, since the fluid has to be displaced less due to smaller axial velocity difference across the transition.
\Fref{fig-nozzle-exit-profiles}(d) quantifies this observation by comparison of the maximum radial flow velocity at the exit with the average axial flow.
Combining the axial and radial flows, streamlines are computed in order to visualize the fluid motion in the stationary state.
As can be seen in the overlaying lines in \fref{fig-nozzle-exit-profiles}(a) and (b), the streamlines show very similar elongational behavior for all $ \alpha $ at the nozzle exit due to the simultaneous decrease of the axial and the radial flow component.
They are, however, not exactly equal, since the maximum axial and radial velocities scale slightly differently with $\alpha$.
Finally, comparing the ratio of axial and radial velocities, we find that the maximum radial flow velocity is always about $\SI{10}{\percent}$ of the maximum axial flow velocity, roughly independent of $ \alpha $.
%\SJM{technical details:}
%\begin{itemize}
%	\item espresso simulations in nozzle, input avg velocity or pressure gradient from python tool
%	\item calculate velocity.vtk in cylindrical coordinates (\TODO{save perl tool}), cyl\_velocity.vtk
%	\item extract slices at $x=(80-120), (200-202), (280-320)$ (inside nozzle, transition averaged over only a few slices, plug flow)
%	\item extract stress along $x$ for different $r$, $r=0$ as reference stress
%	\item extract quantities of interest from slices (gnuplot stats\_max)
%\end{itemize}
\\
The fluid stress along the axial direction for different offsets is shown in \fref{fig-nozzle-flow-deriv-decomp-newtonian} for $\alpha=0$ and $\alpha=0.75$.
In addition to the total fluid stress, we plot the shear and elongational component separately.
To do so, we first decompose the rate of strain tensor into the shear and elongational components
\begin{align}
\label{eq-nozzle-fluidstress-decomposition-2}
\dot{S}_{ij} = \dot{S}_{ij}^{\mathrm{shear}} + \dot{S}_{ij}^{\mathrm{elong}} \eqspace ,
\end{align}
where $\dot{S}^{\mathrm{elong}}$ is a diagonal tensor and $\dot{S}^{\mathrm{shear}}$ contains only off-diagonal elements.
Using this decomposition --- further details can be found in \sisref{3} ---, we can define the shear and elongational components of the fluid stress as
\begin{align}
\label{eq-nozzle-fluidstress-decomposition-shear}
%	\fluidstress^\mathrm{shear} \defeq \dynvisc\qty(\rateofstrain)\sqrt{4\straintensor_{xy}^2}
\fluidstress^\mathrm{shear} \defeq \dynvisc\qty(\rateofstrain)\sqrt{4\straintensor_{xr}^2}
\end{align}
and
\begin{align}
\label{eq-nozzle-fluidstress-decomposition-elong}
%	\fluidstress^\mathrm{elong} \defeq \dynvisc\qty(\rateofstrain) \sqrt{2(\straintensor_{xx}^2 + \straintensor_{yy}^2 + \straintensor_{zz}^2)} \eqspace .
\fluidstress^\mathrm{elong} \defeq \dynvisc\qty(\rateofstrain) \sqrt{2(\straintensor_{xx}^2 + \straintensor_{rr}^2 + \straintensor_{\theta \theta}^2)} \eqspace .
\end{align}
%Note that $\straintensor_{xz}=\straintensor_{yz}=0$ in the $x$-$y$-plane.
Note that $\straintensor_{x \theta}=\straintensor_{r \theta }=0$, since no azimuthal flow components are present, but that nevertheless $ \straintensor _{ \theta \theta  } \neq 0$ as detailed in \sisref{3}.
Thus, the total fluid stress is obtained from \eqref{eq-nozzle-fluidstress-decomposition-shear} and \eqref{eq-nozzle-fluidstress-decomposition-elong} via:
\begin{align}
\label{eq-nozzle-fluidstress-decomposition-1}
\fluidstress & = \sqrt{ \qty(\fluidstress^\mathrm{shear\vphantom{g}})^2 + \qty(\fluidstress^\mathrm{elong})^2 }
\end{align}
Along the channel center (cf. \fref{fig-nozzle-flow-deriv-decomp-newtonian}(a) and (e)), all shear components of the stress vanish, leaving only the elongational ones, which show a clear peak at the exit.
Considering the symmetry, this peak is caused solely by the axial flow deceleration.
\\
With increasing radial offset from the center, as can be seen in \fref{fig-nozzle-flow-deriv-decomp-newtonian}(b-d and f-h) for offsets $1.5\cellradius$, $3.0\cellradius$, and $4.5\cellradius$, the influence of the shear components increases significantly.
It can also be seen that the peak of the fluid stress is not only determined by the elongational flow components, but also partly by the shear component $\straintensor_{xr}=\half \qty(\pdv{\velocity_x}{r} + \pdv{\velocity_r}{x})$.
This is further discussed in \sisref{2} in the SI.
The radial offset at which the shear stress inside the nozzle channel exceeds the magnitude of the fluid stress peak depends on the shear thinning strength of the bioink:
when comparing \fref{fig-nozzle-flow-deriv-decomp-newtonian}(c) and (g), the stress peak for the Newtonian fluid is already smaller than the fluid stress inside the nozzle channel, while for $\alpha=0.75$ it is still higher.
When selecting shear thinning bioinks in bioprinting, it is thus important to keep in mind that the relative significance of the radial flows at the nozzle exit, both elongational and corresponding shear components, increases when a stronger shear thinning bioink is used.

\subsection{Cell flowing through the nozzle exit}
\label{sec:cell-nozzle}
In this section, we investigate the influence of the flow transition on cells passing the exit of the printer nozzle using our computer simulations and actual micrographs of cells flowing through a real 3D bioprinter nozzle.
As discussed in \sref{sec:flow-analysis}, elongational flow components on a short length scale ($\approx 2 \channelradius$) occur at the nozzle exit.
These act in different ways on the cell, depending on its radial position when passing the transition:
%\\
\subsubsection{Centered cell}
Flowing along the center of the channel, the cell experiences symmetric flow conditions also when passing through the nozzle exit.
The deceleration in flow direction leads to an axial compression, while the radial flow stretches the cell in radial direction, leading to an oblate deformation of the cell.
As can be seen in the simulation snapshots in \fref{fig-nozzle-cellstress-newtonian}(a) and (b) for the centered flowing cell, its stress uniformly increases inside the whole cell volume during this elongational deformation. 
After the transition, the cell quickly relaxes towards its spherical equilibrium shape inside the bioink strand.

Next, we assess the cellular stress resulting from the various flow regimes and ink properties.
As can be seen in \fref{fig-nozzle-cell-stress-offsets-materials}(a), an increase in the shear thinning strength of the bioink leads to a decreasing cell stress inside the nozzle channel, as expected from the experimentally observed increased cell survival in more shear thinning bioinks~\cite{ouyang_effect_2016,billiet_3d_2014}.
In contrast to these beneficial effects of shear thinning inside the nozzle, we find that the importance of the elongational stress peak at the nozzle exit notably increases relative to the stress inside the nozzle when $\alpha$ is increased: for the Newtonian case (dark blue line \fref{fig-nozzle-cell-stress-offsets-materials}(a)), 
cell stress increases by approximately 50\% from $\SI{0.9}{\kilo\pascal}$ to $\SI{1.3}{\kilo\pascal}$ during the transition, 
while for the most shear-thinning bioink (light blue line) it increases six-fold from $\SI{0.1}{\kilo\pascal}$ to $\SI{0.6}{\kilo\pascal}$.
\\
Besides cell stresses, an important measure to assess cell damage is cell strain, see \eg~\cite{fischer_calcium_2022}.
Due to the symmetry at the channel center, we define an axial strain $\alpha_1\defeq l_x/(2\cellradius)$ and a radial strain $\alpha_2\defeq l_r/(2\cellradius)$, as the maximum elongation of the cell in the considered direction divided by the cell's reference diameter.
%We define the strains as the maximum elongation of the cell in the considered direction, normalized by the cell's reference diameter, $\alpha_x = d_x/(2 \cellradius)$ (and similar expressions for $y,z$ or $r$).
As shown in \fref{fig-nozzle-flow-elongation-stressmax-estimate}, the behavior of these cell strains is similar to that of the cell stresses in the paragraph above.
Independent of the shear thinning exponent $\alpha$, the axial strain $ \alpha _1$ of the cell's bullet shape inside the nozzle channel is almost negligible, and only a clear peak in deformation is observed when passing the nozzle exit.
The radial strain $ \alpha _2$, on the other hand, already starts with a significant difference from the equilibrium shape. % \SG{ due to the increased flow curvature of less shear thinning bioinks}. \SG{ why does the flow curvature cause radial deformation? Maybe remove marked part? } \SJM{$\checkmark$, was a mislead explanation attempt}
A cell suspended in a highly shear thinning bioink flowing at the nozzle center therefore experiences only the elongational flow right at the nozzle exit, while remaining almost undeformed otherwise.
\\
\subsubsection{Off-centered cell}
We now observe a cell flowing near the nozzle wall
Here, the elongational flow at the nozzle exit is combined with shear components inside the nozzle.
When passing the transition, the cell is pushed in radial direction leading to a non-ellipsoidal change in shape, before it relaxes towards the equilibrium shape in the bioink strand.
An overall decrease of the cell stress when passing through the transition can be observed in the simulation snapshots for the off-centered flowing cell in \fref{fig-nozzle-cellstress-newtonian}(a) and (b).
Compared to centered cells in \fref{fig-nozzle-cell-stress-offsets-materials}(a), the importance of elongational relative to shear stress decreases for off-centered cells as shown in \fref{fig-nozzle-cell-stress-offsets-materials}(b)-(d).
Indeed, for off-centered cells, the relaxation from the shear-dominated axial flow inside the nozzle channel to the stress-free plug flow in the bioink strand is the most significant effect.
\\
We determine this relaxation time scale $\tau$ for every simulation by fitting an exponentially decaying function to the cell stress versus time data (see SI \sifref{8}).
\Fref{fig-nozzle-cell-stress-offsets-materials}(e) shows the obtained relaxation times for all cell offsets as function of $\alpha$.
We find that the relaxation time increases with increasing shear thinning strength $ \alpha $ when keeping $ \dynvisc_0$ constant.
This is caused by the larger viscosity of the bioinks with higher $\alpha$ for low rates of strain (cf. \fref{fig-viscosity-shearrate-materials}), resulting in a higher resistance of the fluid against the cell shape relaxation.
% \SG{ ok? }
%\BF{So, when you increase alpha, you also increase $\eta_0$? Why would you do that? This is confusing. }
%\BF{Didn't you state earlier that the cell viscosity is the same as the suspension fluid viscosity in your simulations? So if it is true that you increase $\eta_0$ with higher alpha (which I find very confusing), the cell would then automatically also have a higher viscosity, which explains the slower relaxation time. }
%\SJM{We keep $\dynvisc_0$ constant when increasing $\alpha$. What we refer to here is the viscosity difference visible in figure 1(a) for low $\rateofstrain$. With that, both fluid and cell have a larger viscosity compared to the Newtonian case. Also note that the Newtonian reference bioink has a viscosity of $\dynvisc_0/2$.}
%\maybe{There is also a change in relaxation times with the initial radial position of the cell.	The cell flowing in the Newtonian fluid has an increased $\tau$ with increasing distance from the center, however, for $\alpha=0.75$ this order is reversed, with the cell flowing at the center showing the largest relaxation time.} \SJM{or leave that out?}
Similar to our observations of the fluid stress at the nozzle exit in \sref{sec:flow-analysis}, we find in \fref{fig-nozzle-cell-stress-offsets-materials}(a to d) that the cell stress peak at the nozzle exit becomes more significant compared to the cell stress inside the nozzle channel when the cell is closer to the center and for stronger shear thinning fluids.
%We plot the maximum cell stress for all off-centered flowing cells in \fref{fig-nozzle-cell-stress-offsets-materials}(f)\TODO{??}
% experimental observations
%
\subsubsection{Microscopy experiments}
\begin{figure}
	\caption{\label{fig-nozzle-cellstress-newtonian}Internal cell stress distribution of cells flowing at different offsets through the nozzle exit in (a) the Newtonian fluid and (b) the shear thinning bioink with $\alpha=0.75$.
	(c) Experimental image of cells exiting a $\SI{100}{\micro\meter}$ radius nozzle in $\SI{2}{\percent}$ alginate bioink. 
	Left green boxes indicate radially/axially elongated cells flowing in/off-center, respectively. 
	Right green box indicates a cell after relaxation back to equilibrium.
	Movies of both simulation and experiment can be found in the supplementary material.
	}
	\centering
	\includegraphics[width=\linewidth]{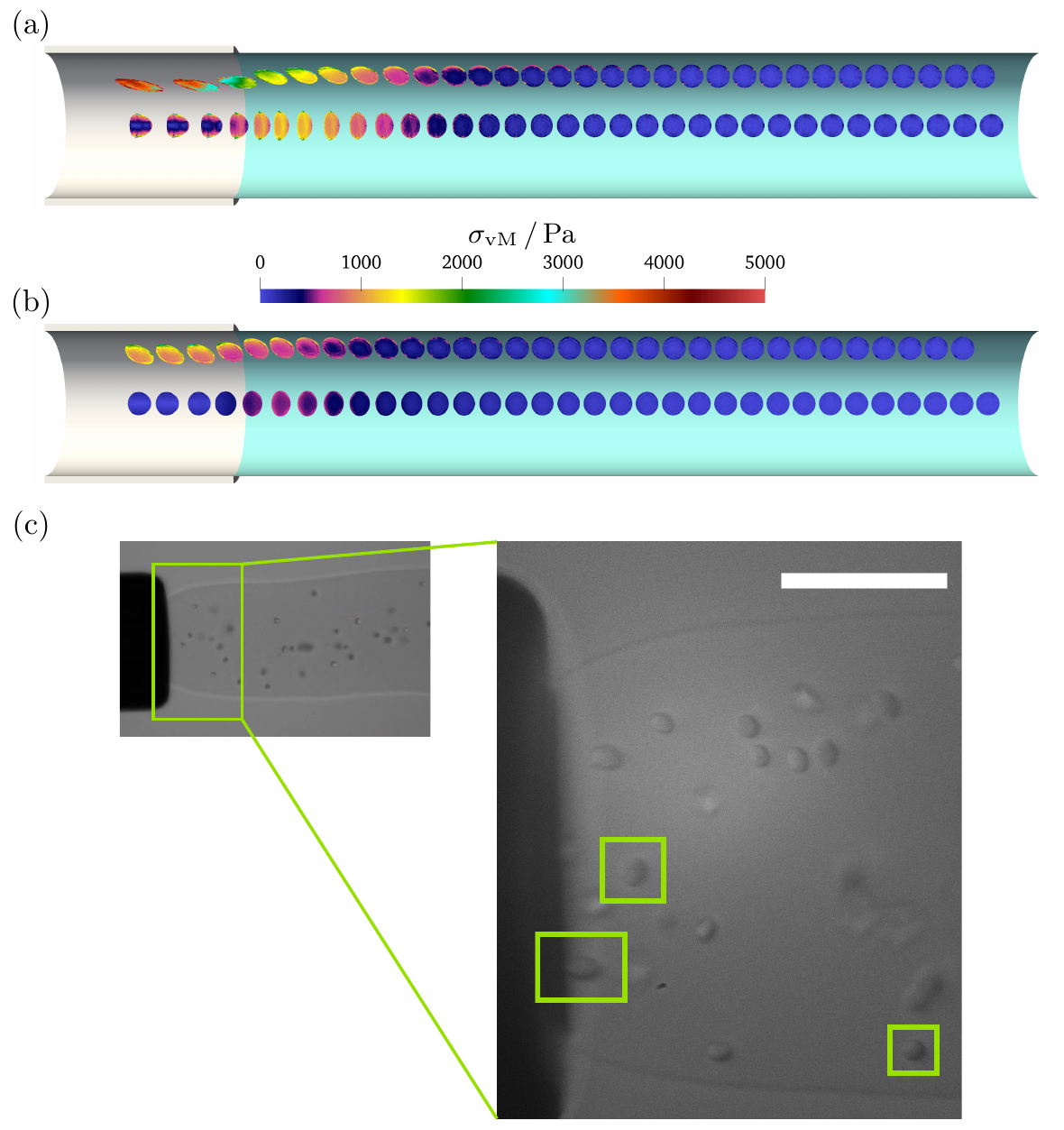}
\end{figure}

\begin{figure}
	\centering
	\caption{\label{fig-nozzle-cell-stress-offsets-materials}Change of the cell stress when passing through the nozzle exit and flowing in the free bioink strand for increasingly shear thinning bioinks. From (a) to (d), the data is given for the initial cell's radial offsets $0$, $1.5\cellradius$, $3\cellradius$, and $4.5\cellradius$. (e)~Relaxation times $\tau$ of the cell stress when flowing in the free bioink strand as function of $\alpha$ and the initial radial cell offset. 
	}
	% zplot-nozzle-cell-stress.plt, zplot-nozzle-cell-relaxationtimes.plt, 
	\includegraphics[width=\linewidth]{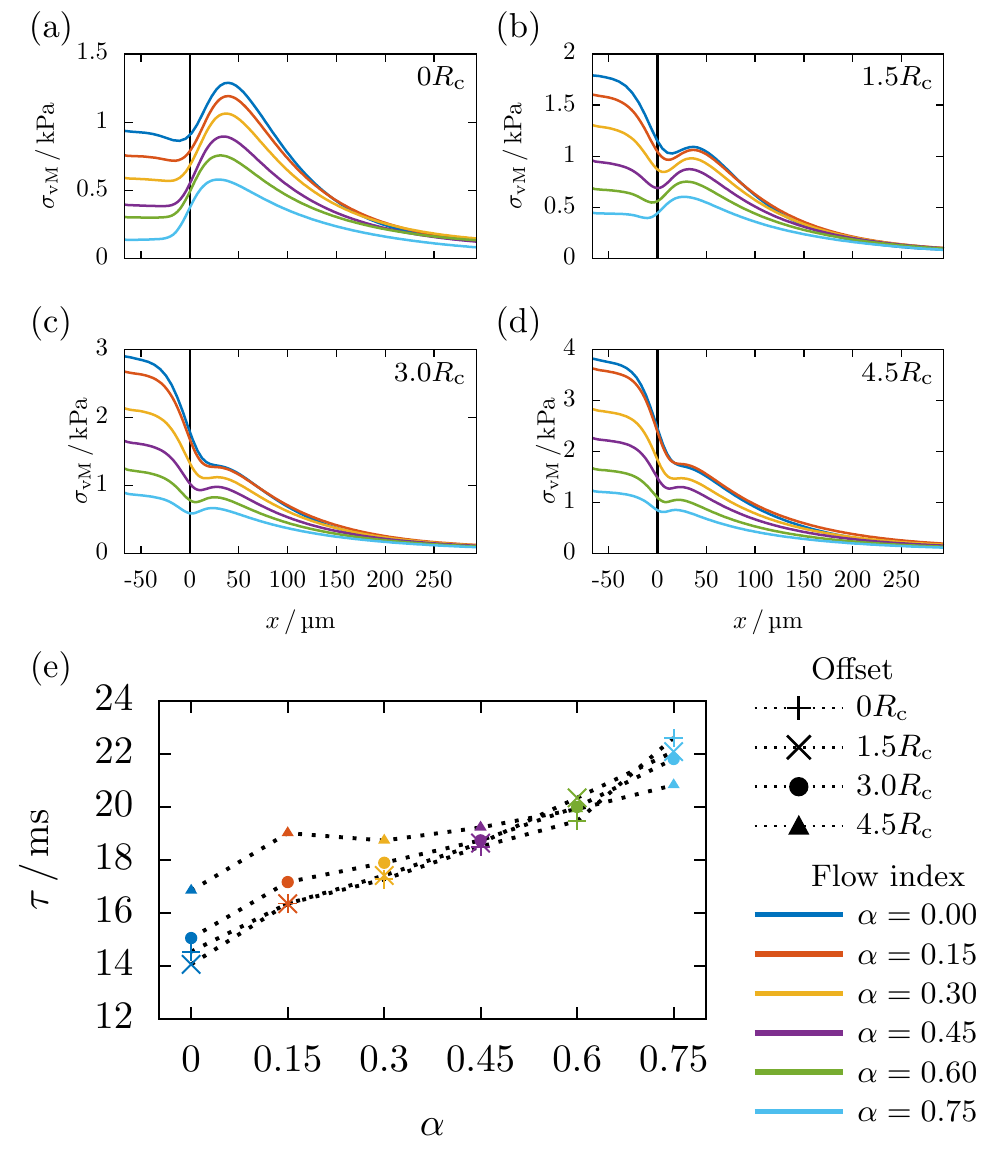}
\end{figure}

\begin{figure}
	\centering
	\caption{\label{fig-nozzle-flow-elongation-stressmax-estimate}Cell strain for a cell flowing in the center (a) $\alpha_1$ in $x$-direction and (b) $\alpha_2$ in $r$-direction at the nozzle exit for the different bioinks. The colors correspond to the flow index as in \fref{fig-nozzle-cell-stress-offsets-materials}.
	}
	\includegraphics[width=\linewidth]{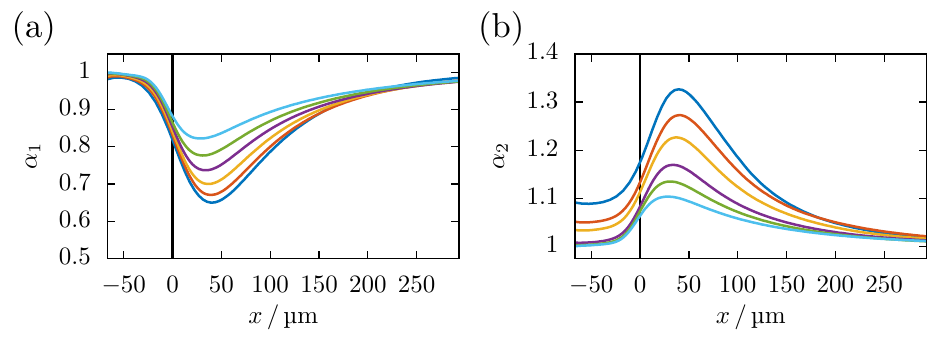}
\end{figure}

To verify our numerical predictions, we image with a high speed camera a bioink strand with cells flowing out of a printing nozzle into a larger reservoir of water.
Details of this imaging setup are included in the SI (cf.~\sisref{1}).
With the objective focused at the tip of the nozzle (inner radius $\SI{100}{\micro\meter}$), the micrograph in \fref{fig-nozzle-cellstress-newtonian}(c) shows cells suspended in a strand of $\SI{2}{\percent}$ alginate bioink during extrusion at a flow rate of $\SI{10}{\micro\litre\per\second}$.
As can be seen in the marked areas in \fref{fig-nozzle-cellstress-newtonian}(c), cells flowing close to the center exhibit a radially elongated change of shape, while cells flowing near the nozzle wall show an axial elongation.
In accordance with our simulations, we observe the cells in the experiment relaxing towards their spherical stress-free shape shortly after the nozzle exit.

\subsection{Prediction of elongational stress, cell stress and cell strain during bioprinting}
\label{sec:prediction}
The methods employed in sections~\ref{sec:cell-cylinder}, \ref{sec:flow-analysis}, and \ref{sec:cell-nozzle} lead to accurate predictions for important parameters such as cell strain or stress, but require numerical simulations with specialized software.
As a practical tool, we develop in the following a simpler yet still accurate method to predict important cell quantities from the printing parameters only.
%These can be used to assess conditions for cell viability in different experimental situations (see the \sref{sec:application} for an example).

\subsubsection{Elongational fluid stress at the nozzle exit}
\label{sec:elongational-fluid-stress-esimtate}
\begin{figure}
	\centering
	\caption{\label{fig-nozzle-cell-dimensions-materials}The average elongational stress $\avgelongstress$ across the nozzle exit from \fref{fig-nozzle-flow-deriv-decomp-newtonian}(a and e) can be estimated from \eqref{eq-elongation-stress-estimate}.}
		% using solely quantities known from the nozzle channel flow with \eqref{eq-elongation-stress-estimate}.}
	\includegraphics[width=0.75\linewidth]{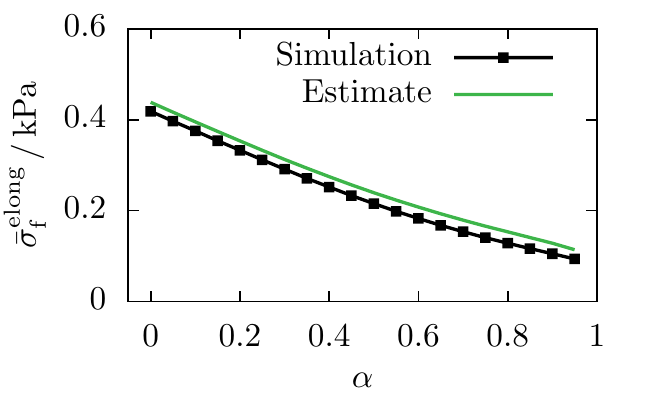}
\end{figure}

%Similar to the shear stress for off-centered cells, we can use the elongational stress at the nozzle exit to predict the cell deformation for centered flowing cells.
To quantify the importance of elongational effects, we define the average elongational fluid stress $\avgelongstress$ which we obtain by averaging $\fluidstress^\mathrm{elong}$ from the simulations along the nozzle axis in an interval of $\mp\channelradius$ around the peak seen in \fref{fig-nozzle-flow-deriv-decomp-newtonian}(a) and (e).
In \fref{fig-nozzle-cell-dimensions-materials} we plot $\avgelongstress$ as function of the shear thinning strength of the fluid.
%From our simulations, we obtain $\avgelongstress$ from the averaged value of $\fluidstress^\mathrm{elong}$ along the nozzle axis, \eg, the average of the peak seen in \fref{fig-nozzle-flow-deriv-decomp-newtonian}(a).
As would be expected from the decreasing pressure gradient, the elongational stress monotonously decreases with $\alpha$.
In order to obliviate the need for full numerical simulations of the entire flow field in practice, we now show that a good estimate for $\avgelongstress$ can be obtained by using a much simpler method for flow field computations~\cite{muller_flow_2020-1}.
%Starting flow calculations inside the nozzle channel, \ie, without the explicit calculation of the changing boundary conditions at the exit, and computationally cheaper using our method from \cite{muller_flow_2020-1}.

For this, we assume that the length of the transition is equal to the nozzle diameter $2\channelradius$, as can be verified by comparing with figure~\ref{fig-nozzle-exit-profiles}(a) and (b) and \fref{fig-nozzle-flow-deriv-decomp-newtonian}(a) and (e).
Starting from the velocity profile of \cite{muller_flow_2020-1}, the change in axial velocity along this length then gives the approximate elongation rate at the nozzle exit:
\begin{align}
\label{eq-elongation-rate-estimate}
%\elongrate = -\pdv{\velocity_x}{x} 
\elongrate & \approx \frac{\velocity_x^\mathrm{max}-\velocity_x^\mathrm{avg}}{2\channelradius}
\end{align}
%We then calculate the stress assuming elongational flow conditions, \ie, $\straintensor_{xx}=-2 \straintensor_{yy}=-2\straintensor_{zz}$, via
Next, we calculate the stress assuming elongational flow conditions, \ie, $\straintensor_{xx}=-2 \straintensor_{rr}=-2\straintensor_{\theta \theta}=-\elongrate$, via
\begin{align}
\label{eq-elongation-stress-estimate}
\avgelongstress = \dynvisc\qty(\sqrt{3}\elongrate) \sqrt{3}\elongrate
\end{align}
which is derived in the SI (cf.~\sisref{3}).
This approximated average elongational stress is in very good agreement with the full numerical simulation of the nozzle exit, as shown in \fref{fig-nozzle-cell-dimensions-materials}. 
\\
%We use this approximation to further estimate the elongational cell stress and strain for centered flowing cells at the nozzle exit in \sref{sec:cell-stress-strain-estimates}.
We use this approximation to further estimate the elongational cell strain and stress for centered flowing cells at the nozzle exit in the next section. % as well as our application example in~\sref{sec:application}.
\subsubsection{Cell stress and strain for centered cells}
\label{sec:cell-stress-strain-estimates}
%To estimate the elongational cell stress, we proceed the following way:
%We use our estimate for the average elongational stress at the nozzle exit from \sref{sec:elongational-fluid-stress-esimtate} together with the theories of Jeffery and Roscoe~\cite{jeffery_motion_1922,roscoe_rheology_1967} for a cell in a stationary elongational flow (cf.~\sref{sec:roscoe-elongational-flow}).
We proceed with an estimation of the maximum stress and strain experienced by cells while flowing inside the nozzle as well as during their transition into the free strand at the nozzle exit.

Starting with the latter, we focus on cells flowing at or close to the nozzle center where (as we have shown in \fref{fig-nozzle-flow-deriv-decomp-newtonian}(a), (b), (e), and (f) above) elongational stresses are the most significant fluid stress contribution.
The theories of Jeffery and Roscoe~\cite{jeffery_motion_1922,roscoe_rheology_1967} contain a solution for the cell strains $ \alpha _1 $ and $ \alpha _2$ in a stationary elongational flow (cf.~\sisref{4 B}).
%\SG{ include the equation here. this is meant to be a simple estimate, so users should not be required to read through the theory first.} \SJM{$\checkmark$}
It reads
\begin{align}
\label{si-eq-roscoe-elong-1}
%2 \roscoedynvisc \frac{\roscoeelongationrate}{g_2^{\prime\prime}} & = \shearmodulus \qty(\alpha_1^2 - \frac{1}{\alpha_1})
\frac{2}{\sqrt{3}} \frac{\avgelongstress}{\shearmodulus} & = \qty(\alpha_1^2 - \frac{1}{\alpha_1}) \int\limits_{0}^{\infty}\frac{\lambda \dd{\lambda}}{\qty(\frac{1}{\alpha_1}+\lambda)^2 \qty(\alpha_1^2+\lambda)^{\frac{3}{2}}}
\end{align}
and can be solved numerically for $\alpha_1$ as function of the elongational fluid stress and the cell's shear modulus.
The other cell strains are $\alpha_2 = \alpha_3 = \alpha_1^{-1/2}$ due to symmetry.
Using the elongation rate from \eqref{eq-elongation-stress-estimate} as input value, we compare the theoretical values with the data obtained from the full numerical simulations in \fref{fig-nozzle-cell-stress-dimensions-materials}(a).
We note that the theory slightly, but consistently, overestimates cell strains.
Indeed, since the elongational flow is experienced by the cell for only a short time span while the theory assumes a stationary elongational flow, this overestimation is to be expected.
%\sout{We note that due to the stationarity condition a stable solution can not be found for very high elongational rates.}\SG{ this makes sense, but where is it relevant here? \fref{fig-nozzle-cell-dimensions-materials}(b) (new) covers all $ \alpha $? } \SJM{moved it to SI}
%We can use our estimate for the average elongational stress at the nozzle exit from \fref{fig-nozzle-cell-dimensions-materials}(a) together with the theories of Jeffery and Roscoe~\cite{jeffery_motion_1922,roscoe_rheology_1967} for a cell in a stationary elongational flow (cf.~\sref{sec:roscoe-elongational-flow}) to compute an approximation for the cell stress and strain occurring at the nozzle transition.
Interestingly, and in line with what has already been observed in \fref{fig-cylinder-time-posy-materials}, Roscoe theory yields surprisingly accurate predictions even for highly shear thinning inks.
We again attribute this to the central role of stresses, instead of flow rates, for the cell deformation process in printing nozzles when these are large compared to the radius of the cell.
With our approximation consistently over-estimating the simulated results, it can be considered as practical upper limit for predicting cell survival. % or functionality estimates.

%In \fref{fig-nozzle-cell-stress-dimensions-materials}(a) above we show our estimate for the elongational cell strain as function of the shear thinning strength for an average extrusion velocity of $\SI{5}{\milli\meter\per\second}$, which consistently over-estimates the one we find in our full numerical simulations.
As a consequence of the stationarity condition assumed by Roscoe theory, it would predict unrealistically large cell strains in the case of printing velocities higher than the $0.5$mm/s used in this work.
In reality, however, the flow through the nozzle exit is highly transient and the stationary state is never attained.
To assess nevertheless the effect of printing speed, we perform additional simulations for cell flowing centered through the nozzle at $\SI{1}{\centi\meter\per\second}$ to $\SI{10}{\centi\meter\per\second}$ average extrusion velocity, in order to cover the typical range of 3D bioprinting speeds.
\Fref{fig-strain-higher-velocity}(a) shows the resulting peak cell strains at the exit from full numerical simulations in comparison to our estimate for $0.5$mm/s in \fref{fig-nozzle-cell-stress-dimensions-materials}(a).
It is apparent that a variation of more than one order of magnitude in flow velocity does hardly affect the cell strains, since the higher velocities significantly decrease the time span during which the high elongational stresses are acting on the cell.
Hence, the printing speed does practically not affect the elongational strains occurring during printing.

Based on this estimate for cell strain, we proceed to estimate the corresponding cell stress for centered cells.
For this, the fluid elongational stress from \eqref{eq-elongation-stress-estimate} is fed into the elongational Roscoe theory given by eqs.~\eqref{si-eq-roscoe-elong-1} and \sieqref{52}.
The result is in good agreement with the full numerical simulations as shown in \fref{fig-nozzle-cell-stress-dimensions-materials}(b) for centered cells (green line).

\subsubsection{Cell stress and strain for off-centered cells}

For off-centered cells, we have shown in \fref{fig-nozzle-cell-stress-offsets-materials}(d) that shear components inside the nozzle are an important contribution to the overall cell stress, especially inside less shear thinning bioinks, where they substantially exceed the stress caused by elongational flows at the nozzle exit.
We next estimate this overall maximum cell strain and stress.

Due to their almost ellipsoidal shape, we choose as strain measure for the off-centered flowing cells now the ellipsoid's major and minor semi-axis $\alpha_1^\prime$ and $\alpha_2^\prime$, which are obtained through computing the equivalent ellipsoid from the deformed cell's inertia tensor, as detailed in~\cite{muller_hyperelastic_2020-2}.

\begin{figure}
	\caption{\label{fig-nozzle-cell-stress-dimensions-materials}(a)~The peak strain and (b)~cell stress for centered flowing cells at the transition can be approximated using our estimate of the average elongational fluid stress from \eqref{eq-elongation-stress-estimate} and the Jeffery and Roscoe theories for a cell in an elongational flow from \sisref{4 B}. For off-centered cells, our flow calculations inside the nozzle channel and the theory of Roscoe for a cell in shear flow. 
	}
	\includegraphics[width=\linewidth]{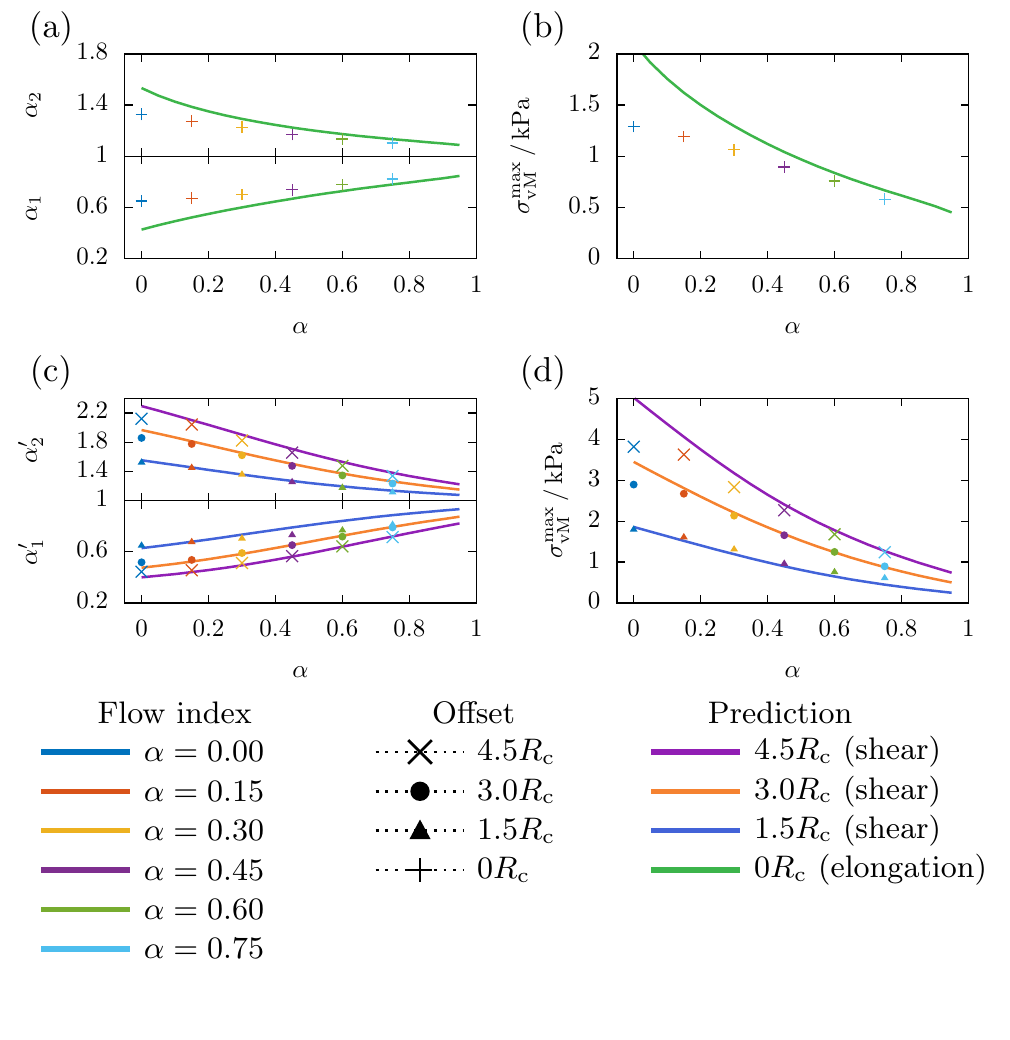}
%	\SJM{below is the same plot with averaged and maximum (offset 4.5) quantities. I think the upper plots look nicer, as the actual simulated data is also matching the corresponding prediction better. } \SG{ yes, remove the lower plot } 	\includegraphics[width=\linewidth]{fig15-avg}
\end{figure}

Starting from the fluid shear stress obtained from our earlier work \cite{muller_flow_2020-1}, we employ the shear part of Roscoe theory in \sieqref{45} and \sieqref{47} and plot the resulting stresses and strains for cells starting at $1.5\cellradius$, $3\cellradius$, and $4.5\cellradius$ in \fref{fig-nozzle-cell-stress-dimensions-materials}(c,d).
Again, we observe very good agreement with the simulations from \sref{sec:cell-nozzle}(ii).

\begin{figure}
	\caption{\label{fig-strain-higher-velocity}
		(a)~Peak elongational cell strain for centered flowing cells passing the transition for an average extrusion velocity of $\SI{1}{\centi\meter\per\second}$, $\SI{2}{\centi\meter\per\second}$, $\SI{5}{\centi\meter\per\second}$, and $\SI{10}{\centi\meter\per\second}$ in comparison to the data of \fref{fig-nozzle-cell-stress-dimensions-materials}(a) for $\SI{5}{\milli\meter\per\second}$. (b)~Maximum cell strain for cells flowing off-centered at $4.5\,\cellradius$ for increasing velocities is in accordance with the prediction of the Roscoe theory.
%		\SG{ remove dotted line to match with figure 13. } \SJM{$\checkmark$}
	}
	\includegraphics[width=\linewidth]{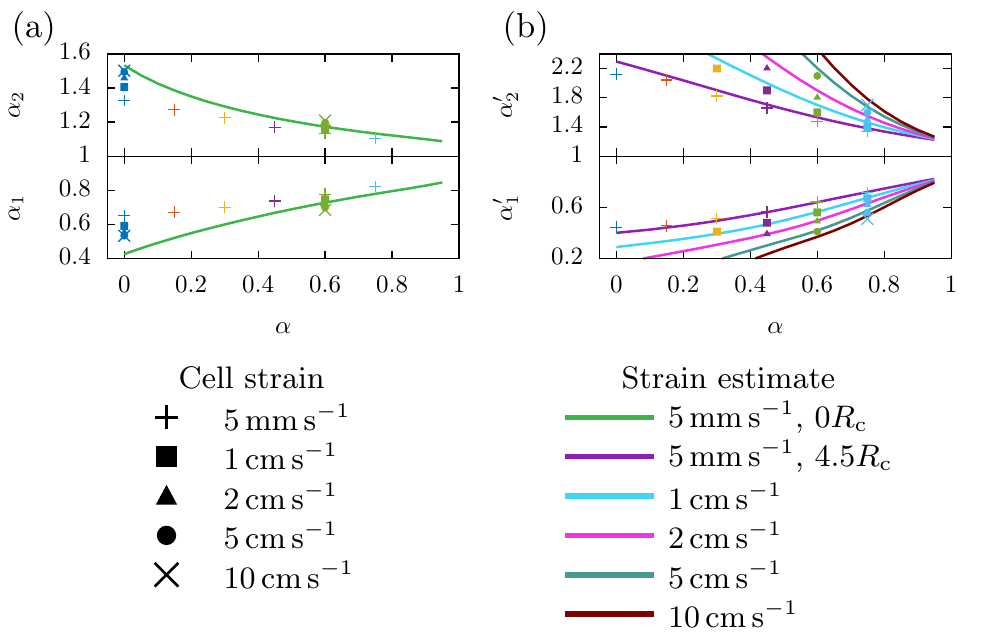}
\end{figure}

Upon increasing the average flow velocity by more than one order of magnitude in \fref{fig-strain-higher-velocity}(b), we find that cells flowing at maximum radial offset  in Newtonian bioinks are not able to attain a stable state while flowing inside the nozzle channel.
However, this limitation is solely a result of the large viscosity of the hypothetical Newtonian fluid, and would not affect a real printing process.
With increasing shear thinning strength, as shown in \fref{fig-strain-higher-velocity}(b), a stable cell shape can be achieved also for high flow velocities of $\SI{10}{\centi\meter\per\second}$.
The maximum cell strains are accurately predicted by Roscoe theory.

% main plot, cell stress and strain prediction
% elong stress higher velocities
\section{Conclusion}
In this work, we investigated the cell stress and strain and the bioink flow behavior during a 3D bioprinting extrusion process using Lattice-Boltzmann numerical simulations together with corresponding qualitative experiments.
The two scenarios considered were the flow inside the nozzle channel as well as at the nozzle exit, where the flow transitions into the free bioink strand.
\\
During the first stage of the printing process while cells are flowing inside the printing nozzle, our simulations showed a bullet-like deformation for cells in the center of the channel and an ellipsoidal shape for cells flowing off-center.
The latter can be understood on the basis of the classical theory of Roscoe~\cite{roscoe_rheology_1967} which relates cell stress to the local fluid stress.
Interestingly, our simulations demonstrate that these relations hold even in realistic shear thinning bioinks, even though they were originally designed for Newtonian fluids only.
The radially inward-directed migration of the  cell due to the shear forces was also found to be independent of the shear thinning strength and solely dependent on the printing pressure.
We show that, when bioprinting at constant flow rate (or velocity), the shear thinning properties reduce the overall cell stress and strain significantly, while this will not be the case for printing processes performed at constant printing pressure.
\\
In the second stage, cells transition into the free printing strand as they exit the printer nozzle.
During this transition, cells are exposed to an elongational flow pattern.
%While a radial deformation also occurs for cells flowing off-center, we find that the shear deformations inside the nozzle are the dominant deformation process in this case.
While a radial deformation also occurs for cells flowing off-center, we find that the shear deformations dominate in this case.
For cells in the channel center, however, this flow causes notable radial stretch of the cells as predicted by our numerical simulations, in qualitative agreement with experimental microscopy images.
We show that this effect becomes particularly relevant for cells flowing in highly shear thinning bioinks, as the shear deformation inside the nozzle can virtually be eliminated, while the radial elongation inevitably takes place (\fref{fig-nozzle-cell-stress-offsets-materials}a).
In addition, we find that the elongational cell strain is practically independent of the extrusion velocity of the bioink, since the faster velocity balances the high elongational stress by reducing the application time. 
The relaxation times of the elongated cells even increase with the shear thinning strength, thus prolonging the time that they remain under strain with potentially harmful side effects (\fref{fig-nozzle-cell-stress-offsets-materials}e).
\\
Using our numerical simulation techniques as a starting point together with the velocity profiles derived in our earlier work~\cite{muller_flow_2020-1}, we finally developed simple estimates for cell stress and/or cell strain for centered as well as off-centered cells. 
%In the nozzle center, the elongational fluid stress during the transition can be accurately estimated from the nozzle radius and maximum flow velocity in good agreement with the full numerical simulation.
%For cells flowing off-center, we derived an estimate for the maximum cell stress -- which occurs inside the printing nozzle and is independent of the bioink rheology -- solely from the printing parameters using the theory of Roscoe\cite{roscoe_rheology_1967} and our flow calculations~\cite{muller_flow_2020-1}.
\\

\begin{acknowledgments}
We thank Nico Schwarm for preparing the cell/alginate sample and help with the imaging experiment.
Funded by the Deutsche Forschungsgemeinschaft (DFG, German Research Foundation) --- Project number 326998133 --- TRR 225 ``Biofabrication'' (subprojects B07 and A01). 
We gratefully acknowledge computing time provided by the SuperMUC system of the Leibniz Rechenzentrum, Garching.
We further acknowledge support through the computational resources provided by the Bavarian Polymer Institute.
\end{acknowledgments}
\bibliography{literature.bib}
\cleardoublepage


\section*{Supplementary material}

# Supplementary Information

## Predicting cell stress and strain during extrusion 3D bioprinting

Sebastian J. Müller[1] , Ben Fabry[2] , and Stephan Gekle[1]

[1] Biofluid Simulation and Modeling, Theoretische Physik VI, Universität Bayreuth, 95440 Bayreuth, Germany (www.gekle.physik.uni-bayreuth.de)

[2] Department of Physics, Friedrich-Alexander University Erlangen-Nürnberg, 91054 Erlangen, Germany

## S-1. BIOPRINTER IMAGING SETUP

MDA-MB-231 breast carcinoma cells were suspended in a 2 % alginate-DMEM solution (sodium alginate PH176, batch nr. 4503283839, JRS Pharma GmbH, Rosenberg, Germany) at a concentration of $10^6 \frac{\text{cells}}{\text{ml}}$. The cell-alginate suspension was then extruded through a stainless steel needle with an inner diameter of 200 µm and a length of 12.7 mm (Nordson EFD, East Providence, USA) at a constant flow rate of $10\,\mu\text{l}\,\text{s}^{-1}$ using a volume-controlled 3-D printer [1]. The needle was dipped into a transparent plastic cuvette (rotilabo, Roth, Germany) filled with PBS solution. The cells were imaged through a non-infinity corrected 10× 0.25 NA objective (Zeiss, Germany) and a lens-less 150 mm tube (Thorlabs, Germany) using a CMOS camera (acA720-520um, Basler, Germany) at an exposure time of 30 µs and a frame rate of 100 Hz.

FIG. S-1. Decomposition of the shear component of the strain rate tensor for (a) the Newtonian fluid and (b) the bioink with $\alpha = 0.75$. In both cases, the local peak of the rate of strain is due to the radial shear component $\frac{\partial u_r}{\partial x}$, while the axial shear decreases monotonously.

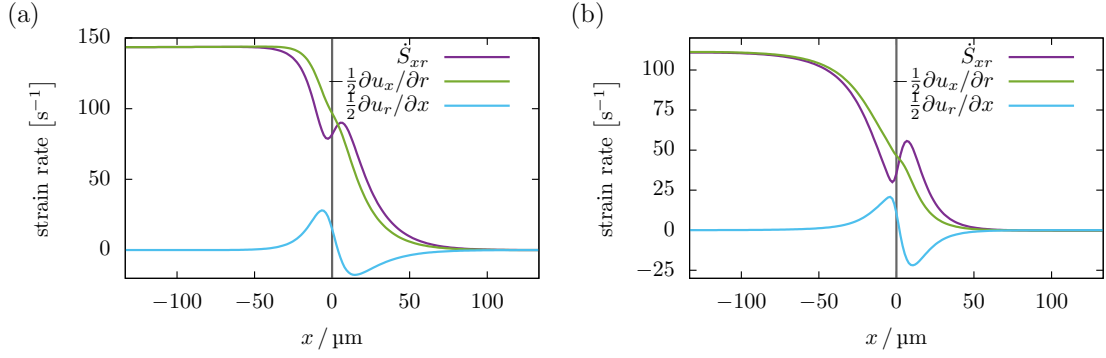

## S-2. FLUID SHEAR STRESS AT THE NOZZLE EXIT

As shown in figure 7 of the manuscript, the total fluid shear stress $\sigma_\mathrm{f}$ along the axial direction always has a local or global peak right after the nozzle exit. When considering the flow close to the channel axis, this peak is clearly a result of the elongational flow at the nozzle exit. Moving closer to the wall, the influence of the elongational components decreases, however, the peak is still present and part of the shear components of the flow. In figure S-1 we decompose the strain rate tensor element $\dot{S}_{xr} = \frac{1}{2}\left(\frac{\partial u_x}{\partial r} + \frac{\partial u_r}{\partial x}\right)$ into its axial and radial shear component. While the axial strain rate $\frac{\partial u_x}{\partial r}$ monotonously decreases along the nozzle exit, the radial strain rate $\frac{\partial u_r}{\partial x}$ increases, changes its sign, and relaxes to zero again due to the localized radial flows at the nozzle exit.

## S-3. FLUID STRESS TENSOR DECOMPOSITION

The components of the rate of strain tensor in a cylindrical coordinate system are given by:

$$\dot{S}_{xx} = \frac{\partial u_x}{\partial x} \tag{S-1}$$

$$\dot{S}_{xr} = \dot{S}_{rx} = \frac{1}{2}\left(\frac{\partial u_r}{\partial x} + \frac{\partial u_x}{\partial r}\right) \tag{S-2}$$

$$\dot{S}_{rr} = \frac{\partial u_r}{\partial r} \tag{S-3}$$

$$\dot{S}_{x\theta} = \dot{S}_{\theta x} = \frac{1}{2}\left(\frac{\partial u_\theta}{\partial x} + \frac{1}{r}\frac{\partial u_x}{\partial \theta}\right) \tag{S-4}$$

$$\dot{S}_{r\theta} = \dot{S}_{\theta r} = \frac{1}{2}\left(\frac{1}{r}\frac{\partial u_r}{\partial \theta} - \frac{u_\theta}{r} + \frac{\partial u_\theta}{\partial r}\right) \tag{S-5}$$

$$\dot{S}_{\theta\theta} = \frac{1}{r}\left(\frac{\partial u_\theta}{\partial \theta} + u_r\right) \tag{S-6}$$

For the present axisymmetric situation, this simplifies to

$$\dot{S}_{ij} = \dot{S}_{ij}^{\text{shear}} + \dot{S}_{ij}^{\text{elong}}, \tag{S-7}$$

$$\dot{S}^{\text{shear}} = \begin{pmatrix} 0 & \frac{1}{2}\left(\frac{\partial u_r}{\partial x} + \frac{\partial u_x}{\partial r}\right) & 0 \\ \frac{1}{2}\left(\frac{\partial u_r}{\partial x} + \frac{\partial u_x}{\partial r}\right) & 0 & 0 \\ 0 & 0 & 0 \end{pmatrix}, \tag{S-8}$$

$$\dot{S}^{\text{elong}} = \begin{pmatrix} \frac{\partial u_x}{\partial x} & 0 & 0 \\ 0 & \frac{\partial u_r}{\partial r} & 0 \\ 0 & 0 & \frac{u_r}{r} \end{pmatrix}. \tag{S-9}$$

We can thus compute the scalar shear rate

$$\left|\dot{S}^{\text{shear}}\right| = \sqrt{2\dot{S}_{ij}^{\text{shear}}\dot{S}_{ij}^{\text{shear}}} = \sqrt{4\dot{S}_{xr}^2} \tag{S-10}$$

$$= \left|\frac{\partial u_r}{\partial x} + \frac{\partial u_x}{\partial r}\right| \tag{S-11}$$

and the scalar elongation rate

$$\left|\dot{S}^{\text{elong}}\right| = \sqrt{2\dot{S}_{ij}^{\text{elong}}\dot{S}_{ij}^{\text{elong}}} \tag{S-12}$$

$$= \sqrt{2(\dot{S}_{xx}^2 + \dot{S}_{rr}^2 + \dot{S}_{\theta\theta}^2)} \tag{S-13}$$

$$= \sqrt{2\left(\frac{\partial u_x}{\partial x}\right)^2 + 2\left(\frac{\partial u_r}{\partial r}\right)^2 + 2\left(\frac{u_r}{r}\right)^2}, \tag{S-14}$$

and the rate of strain

$$\left|\dot{S}\right| = \sqrt{\left|\dot{S}^{\text{shear}}\right|^2 + \left|\dot{S}^{\text{elong}}\right|^2}. \tag{S-15}$$

Using the shear and elongation rates, we define the fluid's scalar shear and elongational stress via

$$\sigma_{\text{f}}^{\text{shear}} := \eta\!\left(|\dot{S}|\right)\left|\dot{S}^{\text{shear}}\right| \tag{S-16}$$

and

$$\sigma_{\text{f}}^{\text{elong}} := \eta\!\left(|\dot{S}|\right)\left|\dot{S}^{\text{elong}}\right|. \tag{S-17}$$

The total fluid stress is thus obtained as:

$$\sigma_{\text{f}} = \eta\!\left(|\dot{S}|\right)|\dot{S}| = \sqrt{\left(\sigma_{\text{f}}^{\text{shear}}\right)^2 + \left(\sigma_{\text{f}}^{\text{elong}}\right)^2} \tag{S-18}$$

When assuming a perfect elongational flow, i.e., $u_x = -\dot{\varepsilon} x$, $u_r = \tfrac{1}{2}\dot{\varepsilon} r$, and $u_\theta = 0$, the fluids rate of strain is given via

$$|\dot{S}| = \left|\dot{S}_{ij}^{\text{elong}}\right| = \sqrt{2\dot{\varepsilon}^2 + 2\!\left(\tfrac{1}{2}\dot{\varepsilon}\right)^2 + 2\!\left(\tfrac{1}{2}\dot{\varepsilon}\right)^2} = \sqrt{3}\dot{\varepsilon} \tag{S-19}$$

and thus follows the fluid stress as:

$$\sigma_{\text{f}} = \sigma_{\text{f}}^{\text{elong}} = \eta\!\left(\sqrt{3}\dot{\varepsilon}\right)\sqrt{3}\dot{\varepsilon} \tag{S-20}$$

## S-4. JEFFERY AND ROSCOE THEORY

An analytical theory describing the deformation and stresses of a cell embedded in a linear flow was proposed by Roscoe [2], based on the work of Jeffery [3]. For convenience, we briefly summarize their theoretical approach and the application of the theory for cells in a shear and an elongational flow scenario in this section.

Jeffery [3] originally solved the problem of the motion of a rigid ellipsoidal particle in a linear flow, i. e., the undisturbed fluid velocity can be written as (using the notation of Roscoe [2])

$$v'_i = e'^{(1)}_{ij} x_j - \zeta'_{ij} x_j,\tag{S-21}$$

where the fluid's rate of strain and vorticity are defined by

$$e'^{(1)}_{ij} = \frac{1}{2}\left(\frac{\partial v'_i}{\partial x_j} + \frac{\partial v'_j}{\partial x_i}\right) \quad \text{and} \quad \zeta'_{ij} = \frac{1}{2}\left(\frac{\partial v'_i}{\partial x_j} - \frac{\partial v'_j}{\partial x_i}\right).\tag{S-22}$$

Jeffery [3] derived the fluid stress acting on the surface of a rigid ellipsoidal particle as

$$p'_{ij} = -p_\mathrm{h} \delta_{ij} + \eta_0 A_{ij},\tag{S-23}$$

with an arbitrary hydrostatic pressure $p_\mathrm{h}$ and a deviatoric tensor $A_{ij}$. Roscoe [2] notes that the deviatoric stress can further be divided into two parts,

$$p'_{ij} = -p_\mathrm{h} \delta_{ij} + 2\eta_0 e'^{(1)}_{ij} + \eta_0 \left(A_{ij} - 2e'^{(1)}_{ij}\right),\tag{S-24}$$

one due to the undisturbed flow from (S-21) and one due to the disturbance of the flow induced by the particle presence. The components $A_{ij}$ in a coordinate system coinciding with the ellipsoid axes can be calculated via (remaining components by cyclic change of indices):

$$A_{11} = \frac{4}{3}\frac{2g''_1 e'^{(1)}_{11} - g''_2 e'^{(1)}_{22} - g''_3 e'^{(1)}_{33}}{g''_2 g''_3 + g''_3 g''_1 + g''_1 g''_2}\tag{S-25}$$

$$A_{12} = \frac{g_1 e'^{(1)}_{12} - \alpha_2^2 g'_3 \zeta'_{12}}{2g'_3(\alpha_1^2 g_1 + \alpha 2^2 g_2)}\tag{S-26}$$

$g_i$, $g'_i$, and $g''_i$ are integrals of the type

$$g_1 = \int_0^\infty \frac{\mathrm{d}\lambda}{(\alpha_1^2 + \lambda)\Delta} \tag{S-27}$$

$$g'_1 = \int_0^\infty \frac{\mathrm{d}\lambda}{(\alpha_2^2 + \lambda)(\alpha_3^2 + \lambda)\Delta} = \frac{g_3 - g_2}{\alpha_2^2 - \alpha_3^2} \tag{S-28}$$

$$g''_1 = \int_0^\infty \frac{\lambda \, \mathrm{d}\lambda}{(\alpha_2^2 + \lambda)(\alpha_3^2 + \lambda)\Delta} = \frac{\alpha_2^2 g_2 - \alpha_3^2 g_3}{\alpha_2^2 - \alpha_3^2}, \tag{S-29}$$

where $\Delta = \sqrt{(\alpha_1^2 + \lambda)(\alpha_2^2 + \lambda)(\alpha_3^2 + \lambda)}$.

Equation (S-23) can directly be employed to compute the stresses acting on a rigid ellipsoid suspended in the undisturbed flow given by (S-21), e.g., a cell inside an elongational flow, as detailed in section S-4 B. Roscoe [2] extended the theory of Jeffery [3] to compute the stresses acting on a non-rigid ellipsoid with moving boundaries, i.e., tank-treading motion. The ellipsoid's surface motion is assumed to be linear — similar to (S-21) — given by

$$v_i = \bar{e}_{ij}^{(1)} x_j - \bar{\zeta}_{ij} x_j, \tag{S-30}$$

with $\bar{e}_{ij}^{(1)}$ and $\bar{\zeta}_{ij}$ denoting respectively the average rate of strain and vorticity inside the particle, which are always equal to their values at the particle surface [2]. The velocity disturbance $\Delta v'_i = v_i - v'_i$ at the particle surface induced by the surface motion of the non-rigid particle is equal to the velocity disturbance of a rigid particle in an undisturbed flow given by

$$v''_i = v'_i - v_i. \tag{S-31}$$

Therefore, (S-24) can be employed to compute the fluid stresses for a non-rigid particle with a moving boundary by simply computing the stress contribution due to the disturbance using the equivalent undisturbed flow (S-31), while keeping the contribution due to the actual undisturbed flow (S-21). Thus:

$$p'_{ij} = -p_\mathrm{h}\delta_{ij} + 2\eta_0 e_{ij}^{\prime(1)} + \eta_0\left(A'_{ij} - 2\left(e_{ij}^{\prime(1)} - \bar{e}_{ij}^{(1)}\right)\right) \tag{S-32}$$

$$= -p_\mathrm{h}\delta_{ij} + \eta_0\left(A'_{ij} + 2\bar{e}_{ij}^{(1)}\right) \tag{S-33}$$

Here, the tensor $A'_{ij}$ is computed for an undisturbed flow of the form given in (S-31) instead of (S-21).

## A. Cell stress and strain in shear flow

Roscoe [2] applies (S-32) to compute the motion of a tank-treading ellipsoidal particle in a linear shear flow. The coordinates of a material point of the particle starting at position $(\tilde{x}_1, \tilde{x}_2, \tilde{x}_3)$ following an elliptical trajectory are given by

$$x_1 = \alpha_1(\tilde{x}_1 \cos(\nu t) - \tilde{x}_2 \sin(\nu t)) \tag{S-34}$$

$$x_2 = \alpha_2(\tilde{x}_1 \sin(\nu t) + \tilde{x}_2 \cos(\nu t)) \tag{S-35}$$

$$x_3 = \alpha_3 \tilde{x}_3, \tag{S-36}$$

where $x_1$, $x_2$, and $x_3$, align with the ellipsoid's semi-axes, thus yielding the surface velocity:

$$v_1 = -\frac{\alpha_1}{\alpha_2} \nu x_2 \tag{S-37}$$

$$v_2 = \frac{\alpha_2}{\alpha_1} \nu x_1 \tag{S-38}$$

$$v_3 = 0 \tag{S-39}$$

The surface velocity defines the rate of strain and vorticity from (S-30). A linear shear flow — commonly described in the global coordinate system as $v'_1 = \kappa x_2$, $v'_2 = v'_3 = 0$ with a shear rate $\kappa$ — written in terms of a coordinate system aligned with the ellipsoid's semi-axes through rotation by an angle $\theta$ is given by

$$v'_1 = \kappa\left(x_1 \sin\theta \cos\theta + x_2 \cos^2(\theta)\right) \tag{S-40}$$

$$v'_2 = -\kappa\left(x_1 \sin^2(\theta) + x_2 \sin\theta \cos\theta\right) \tag{S-41}$$

$$v'_3 = 0. \tag{S-42}$$

From that, the undisturbed fluid's rate of strain and vorticity from (S-21), and, together with (S-30), the fluid stress at the particle surface from (S-32) can be computed.

In a stationary state, the fluid stress must be balanced by the cell stress at the particle surface. As mentioned in section III-A of the manuscript, the cell stress consists of an elastic and a viscous part. For the triaxial ellipsoidal deformation described in (S-34), the elastic stress at the particle surface can be computed from (9) assuming an incompressible cell ($J = 1$). With the corresponding deformation gradient tensor given by $F_{ij} = \alpha_i \delta_{ij}$, the non-zero diagonal elements of the Cauchy stress are then found as

$$\sigma_{11} = \frac{\mu}{3}\left(2\alpha_1^2 - \alpha_2^2 - \alpha_3^2\right), \tag{S-43}$$

with similar expressions for $\sigma_{22}$ and $\sigma_{33}$ obtained by cyclic change of indices. The obtained system of two equations of the stress balance is solved by considering only the differences of the principal stresses, which eliminates the hydrostatic pressure:

$$p'_{11} - p'_{22} = \sigma_{11} - \sigma_{22} \tag{S-44}$$

$$\Leftrightarrow 2\eta_0 \kappa \sin(2\theta) \frac{g''_1 + g''_2}{g''_2 g''_3 + g''_3 g''_1 + g''_1 g''_2} = \mu(\alpha_1^2 - \alpha_2^2) \tag{S-45}$$

$$p'_{11} + p'_{22} - 2p'_{33} = \sigma_{11} + \sigma_{22} - 2\sigma_{33} \tag{S-46}$$

$$\Leftrightarrow 2\eta_0 \kappa \sin(2\theta) \frac{g''_1 - g''_2}{g''_2 g''_3 + g''_3 g''_1 + g''_1 g''_2} = \mu\left(\alpha_1^2 + \alpha_2^2 - \frac{2}{\alpha_1^2 \alpha_2^2}\right) \tag{S-47}$$

Note that $\alpha_3 = \frac{1}{\alpha_1 \alpha_2}$ due to the assumed incompressibility. The viscous contribution of the cell stress can be computed directly from its internal fluid motion (S-37)–(S-39) as:

$$\sigma_{12} = \sigma_{21} = 2\eta_1 e'^{(1)}_{12} = -\eta_1 \nu \frac{\alpha_1^2 - \alpha_2^2}{\alpha_1 \alpha_2} \tag{S-48}$$

Through numerical solution of these equations one obtains the cell stresses and strains as well as the tank-treading frequency as function of the undisturbed fluid's shear rate $\kappa$ (cf. Roscoe[2, eq.(80),(41)]).

To compare it with the Roscoe theory, we numerically assess the viscous shear stress inside the cell from our simulations by extracting the Lattice-Boltzmann strain rate tensor field [4, 5] inside the cell using our method from [6] and averaging over the cell volume. In figure S-2 we show how the resulting viscous component of the cell stress as function of the cell's shear modulus in a linear shear flow of constant strain rate $|\dot{S}| = 100\,\text{s}^{-1}$. For low $\mu$ — i.e., for soft cells — the shear stress inside the cells asymptotically approaches the viscous shear stress of the surrounding undisturbed fluid, as it is increasingly stretched and hence more aligned with the flow. Very stiff cells, on the other hand, will remain their undeformed spherical shape. It can be seen that the transition between these two limits in large parts happens in the stiffness range of biological cells at around $100\,\text{Pa}$ to $10\,\text{kPa}$.

FIG. S-2. Viscous shear stress $\sigma_{12}$ inside the cell as function of the cell stiffness in a shear flow with $|\dot{S}| = 100\,\mathrm{s}^{-1}$. The straight line indicates the viscous fluid stress of the surrounding undisturbed flow field.

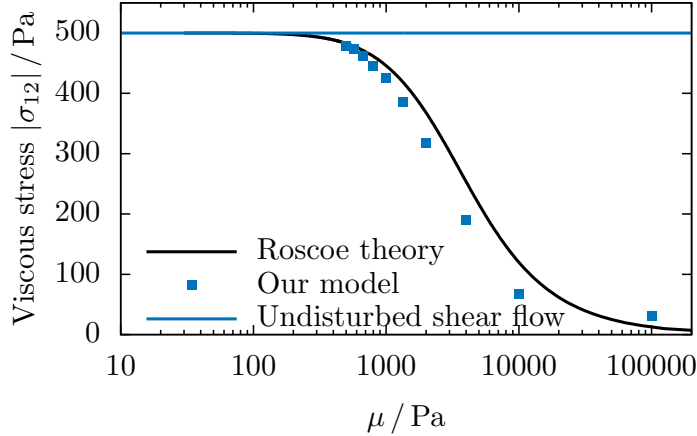

## B.  Cell stress and strain in elongational flow

Roscoe [2] applies (S-23) to compute the steady state deformation and stresses of an ellipsoid in an elongational flow, where the undisturbed velocity is given by

$$v'_1 = \xi x_1 \tag{S-49}$$

$$v'_2 = -\frac{1}{2}\xi x_2 \tag{S-50}$$

$$v'_3 = -\frac{1}{2}\xi x_3 , \tag{S-51}$$

with the elongational rate $\xi$. From this, the rate of strain and vorticity in (S-21) and the deviatoric tensor $A_{ij}$ are calculated. Since the ellipsoid's boundary has no motion, $v_i = 0$. Due to the symmetry of the flow and the incompressibility, $\alpha_2 = \alpha_3 = \alpha_1^{-1/2}$ applies to the ellipsoid. The fluid's normal stress differences from (S-23) are then set to balance the surface stresses of the triaxially elongated particle:

$$p'_{11} - p'_{22} = \sigma_{11} - \sigma_{22} \tag{S-52}$$

$$2\eta_0 \frac{\xi}{g''_2} = \mu\left(\alpha_1^2 - \frac{1}{\alpha_1}\right) \tag{S-53}$$

The numerical solution of this equation yields the cell stresses and strains as function of the undisturbed fluid's elongational rate $\xi$. We note that due to the stationarity condition a stable solution can not be found for very high elongational rates.

9

## S-5. APPLICABILITY OF ROSCOE THEORY FOR SHEAR THINNING BIOINKS

In figure 5(c and d) of the manuscript we show the cell stress as function of the fluid stress for different cell starting positions in the channel in a Newtonian bioink and for the maximum radial offset for increasing shear thinning strength. Figure S-3 shows additional data curves for all investigated bioinks, i.e., data similar to figure 5(c) for different $\alpha$. In Addition to the fluid stress on the lower $x$-axis, the upper $x$-axis gives the radial position of the cell in units of the cell radius.

As mentioned in section III A of the manuscript, the key property determining cell motion is the shear stress. To underline this, we plot in figure S-4 the cell stress data from figure 5(d), but with respect to the rate of strain instead of the shear stress. Due to the similar velocities, the range of the $|\dot{S}|$-axis is similar for all flow indices. Therefore — instead of collapsing onto a master curve as in figure 5(d) — the curves fan out, suggesting a weaker dependency of the cell stress on the shear rate for increasingly shear thinning bioinks. This, however, is slightly misleading, since it neglects the change in viscosity of the surrounding liquid.

In section III E we find that the influence of higher extrusion velocities on the elongational cell strain is almost negligible. However, this does obviously not apply to the shear conditions inside the nozzle channel, as a higher pressure gradient is necessary to produce larger flow velocities. In figure S-5 below, we show data similar to that of figure 5(c and d), for a cell starting at the largest radial offset in a bioink with $\alpha = 0.6$ for average extrusion velocities of $1\,\mathrm{cm\,s^{-1}}$, $2\,\mathrm{cm\,s^{-1}}$, and $5\,\mathrm{cm\,s^{-1}}$, demonstrating the validity of the Roscoe theory also for higher velocities.

FIG. S-3. The cell stress inside the nozzle channel as function of the local shear stress for all used bioinks, as in figure 5(c). The upper $x$-axis gives the radial position of the cell in units of the cell radius.

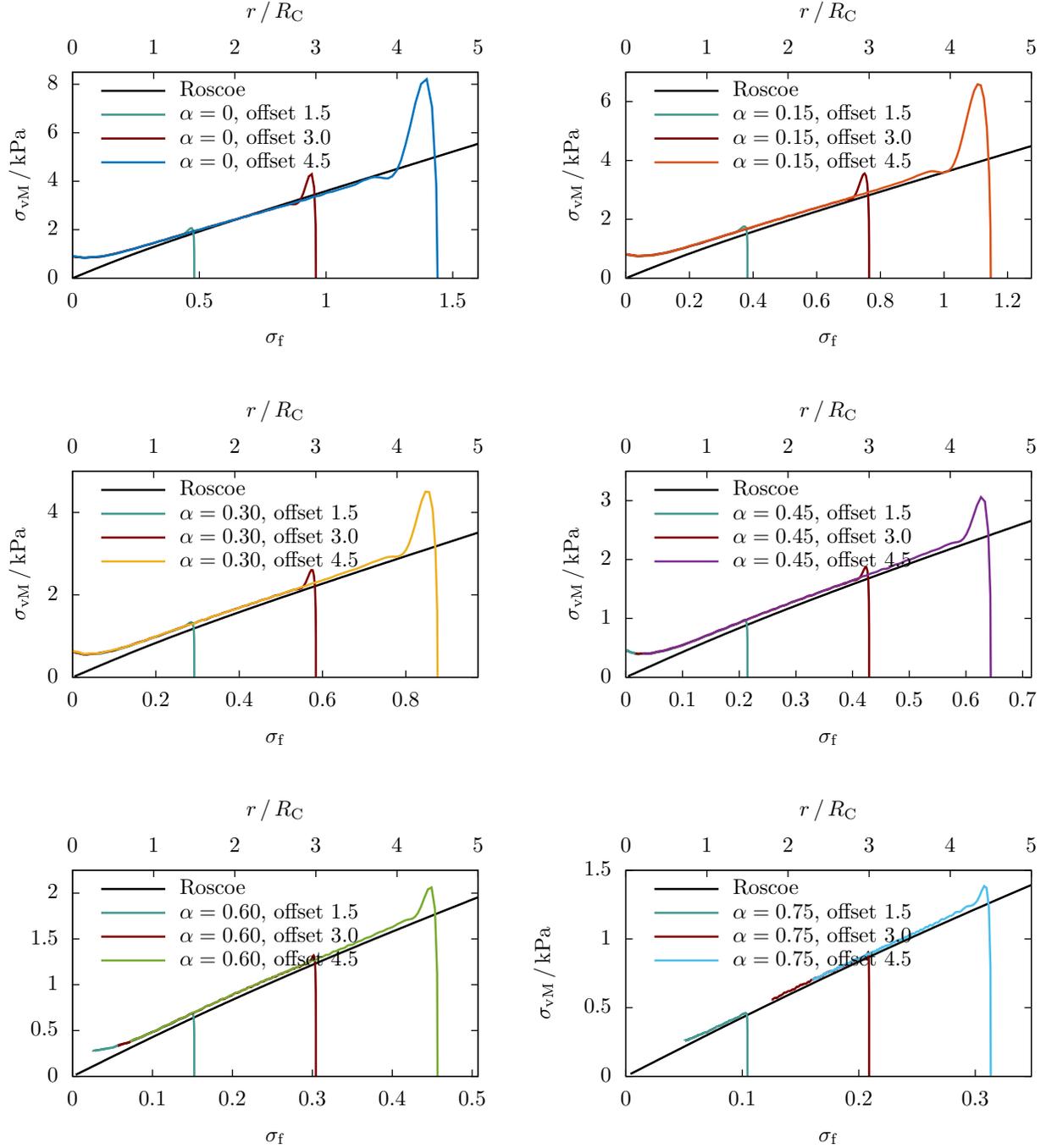

FIG. S-4. Data from figure 5(d), but plotted versus the local rate of strain $|\dot{S}|$ of the fluid. Due to the constant average velocity, the shear rates experienced by the cells are of similar magnitude.

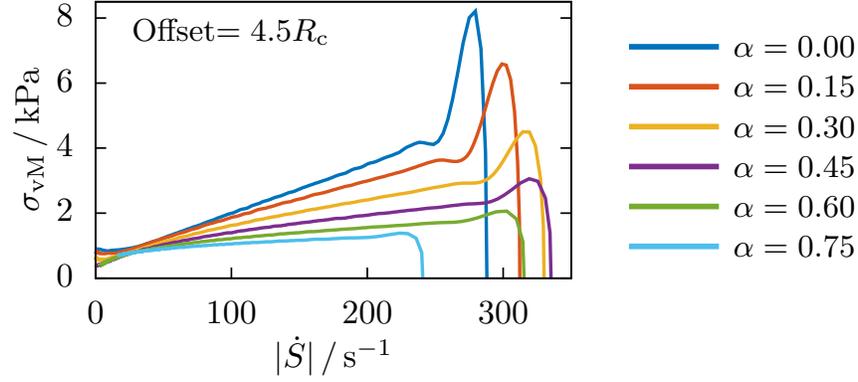

FIG. S-5. The cell stress as function of the fluid stress (similar to figure 5(d) for a cell starting at the largest radial offset in a bioink with $\alpha = 0.6$ for average extrusion velocities of $1\,\mathrm{cm\,s^{-1}}$, $2\,\mathrm{cm\,s^{-1}}$, and $5\,\mathrm{cm\,s^{-1}}$, demonstrating the validity of the Roscoe theory also for higher velocities.)

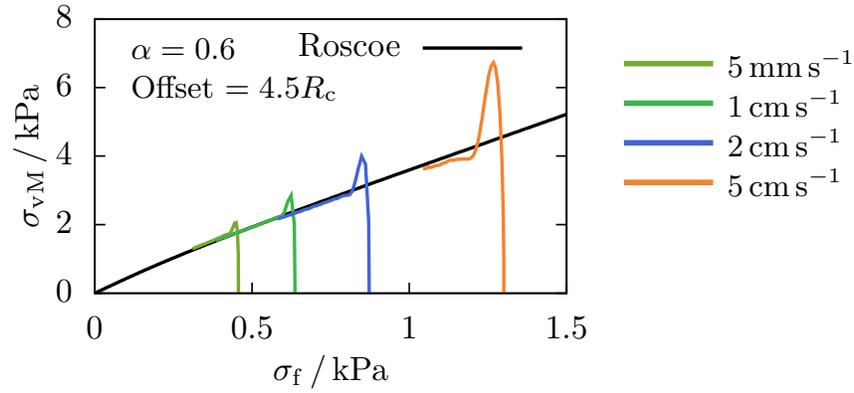

12

## S-6. RIGID SPHERE IN FLOW

To compute the additional stress caused by a rigid sphere in pure shear flow, we start from the strain rate tensor (S-22) of the undisturbed flow. As usual, for a shear rate $\kappa$, this is given by

$$\underline{e}'^{(1)} = \frac{1}{2} \begin{pmatrix} 0 & \kappa & 0 \\ \kappa & 0 & 0 \\ 0 & 0 & 0 \end{pmatrix} \tag{S-54}$$

The sphere is neutrally buouyant as well as force- and torque-free. We then compute the so-called stresslet (see, e. g., [7, eq. (2.32)]) which embodies the additional stress in the fluid due to the presence of the sphere

$$\begin{aligned}\underline{S} &= \frac{20}{3}\pi\eta_0 R_c^3 \underline{e}'^{(1)} \\ &= \frac{20}{6}\pi\eta_0 R_c^3 \kappa \begin{pmatrix} 0 & 1 & 0 \\ 1 & 0 & 0 \\ 0 & 0 & 0 \end{pmatrix}\end{aligned} \tag{S-55}$$

This quantity is normalized by the shear stress of the undisturbed fluid integrated over the sphere volume

$$S_\mathrm{f} = \frac{4}{3}\pi R_c^3 \eta_0 \kappa \tag{S-56}$$

thus leading to the dimensionless stresslet

$$\begin{aligned}\underline{S}^* &= \frac{\underline{S}}{S_\mathrm{f}} \\ &= \frac{5}{2}\begin{pmatrix} 0 & 1 & 0 \\ 1 & 0 & 0 \\ 0 & 0 & 0 \end{pmatrix}.\end{aligned} \tag{S-57}$$

(S-57) is given in the laboratory system. To express it in the body-fixed coordinate system of the cell, we require a rotation by $45°$ (cf. left inset in figure 4) given by the matrix

$$\underline{M}_{45} = \frac{\sqrt{2}}{2}\begin{pmatrix} 1 & -1 & 0 \\ 1 & 1 & 0 \\ 0 & 0 & 1 \end{pmatrix}. \tag{S-58}$$

13

The final result is

$$\underline{S}^*_{\text{rot}} = \underline{M}_{45} \, \underline{S}^* \, \underline{M}_{45}^{\text{T}}$$

$$= \frac{5}{2} \begin{pmatrix} -1 & 0 & 0 \\ 0 & 1 & 0 \\ 0 & 0 & 0 \end{pmatrix} \tag{S-59}$$

thus furnishing an explanation for the cell stress at low flow rates in figure 4.

### S-7. UNIAXIAL STRETCHING OF AN ELASTIC BEAM

In the limit of high Capillary numbers, the elastic components of the cell stress tensor approximately develop according to the ratio $\sigma_{11} : \sigma_{22} : \sigma_{33} = 2 : -1 : -1$. This ratio is equivalent to what would be expected from the uniaxial extension of an elastic beam, as briefly outlined in the following. The uniaxial stretching with a factor $\alpha_1 = a$ in $x_1$-direction of an isotropic, incompressible material results in $\alpha_2 = \alpha_3 = \frac{1}{\sqrt{a}}$ for the remaining principal stretches. The left Cauchy-Green deformation tensor is hence given by $B = \text{diag}\left(a^2, \frac{1}{a}, \frac{1}{a}\right)$, which can be inserted into (S-43) of the manuscript in order to obtain the stress components as:

$$\sigma_{11} = \frac{2}{3}\mu\left(a^2 - \frac{1}{a}\right) \tag{S-60}$$

$$\sigma_{22} = \sigma_{33} = -\frac{1}{3}\mu\left(a^2 - \frac{1}{a}\right) \tag{S-61}$$

## S-8. STRESS RELAXATION INSIDE THE BIOINK STRAND

In figure S-6 we show the individual fits of the cell stress relaxation times $\tau$ from figure 9(e) of the manuscript. As a fit function we use an exponential decay of the form

$$\sigma_{\text{vM}}(t) = \sigma_{\text{vM}}^{\text{arb.offset}} + \sigma_{\text{vM}}^{(0)} \exp\left(-\frac{t-t_0}{\tau}\right), \tag{S-62}$$

where $\sigma_{\text{vM}}^{(0)}$ denotes the cell stress at $t_0$ (indicated by the gray area), and $\sigma_{\text{vM}}^{\text{arb.offset}}$ is an arbitrary small offset. At $t_0$, the cell passes the transition (i.e. $x = 0$).

When the cell relaxes in a quiescent fluid, i.e., when we disable any imposed flow or external pressure, the relaxation times of the cell decrease slightly. This is shown in figure S-7, where we compare the relaxation times of cells inside quiescent fluid to those of cells passing through the nozzle exit into the bioink strand (cf. figure 9(e))

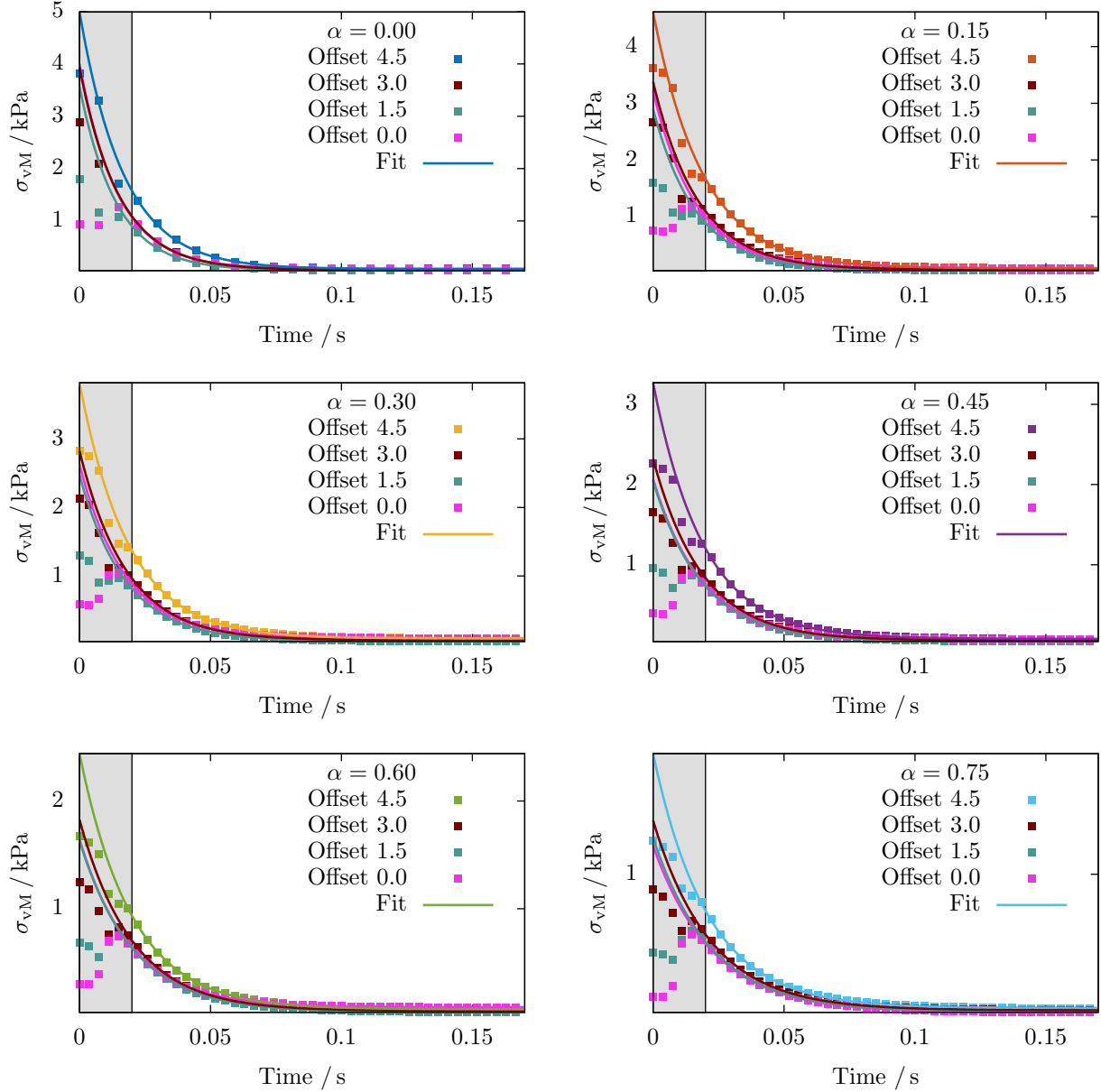

FIG. S-6. Relaxation time fit of the cell stress in the bioink strand using an exponentially decaying function. The fit excludes they gray shaded area.

16

FIG. S-7. Relaxation times of cells suspended in quiescent liquid (solid lines), compared to the data in figure 9(e) (dotted lines).

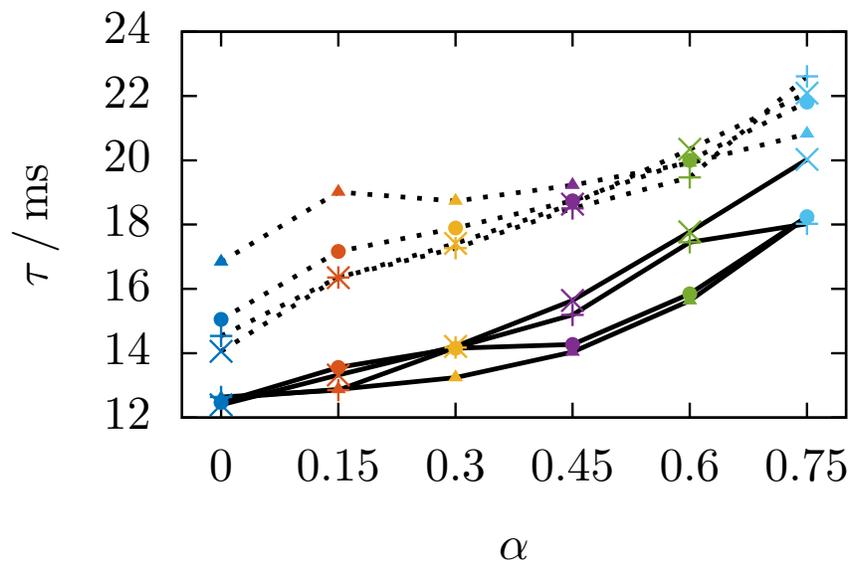

## S-9. CELL STIFFNESS VARIATION

All simulations in the manuscript were performed using a cell with a fixed shear modulus. We plot in figure S-8 the results of figure 9 and figure 10 (gray lines) together with the same data of a softer cell with a shear modulus of $\mu = 500\,\mathrm{Pa}$.

The maximum cell stress in figure S-8(a) to (e), i.e., the peak right after the exit as well as the magnitude inside the nozzle channel before the exit, are approximately half of the value obtained for the stiffer cell. This is due to the stress calculation in (9), where the shear modulus scales the influence of the deformation. Additionally, the cell strain peaks in figure S-8(f) and (g) are of similar order. An inverse scaling with the stiffness is observed for the stress relaxation time $\tau$ in figure S-8(e), showing values about twice as large for the soft cell compared to the stiff one.

FIG. S-8. Influence of the cell stiffness: (a) to (e) data from figure 9 and (f, g) data from figure 10 for a softer cell with $\mu = 500\,\mathrm{Pa}$.

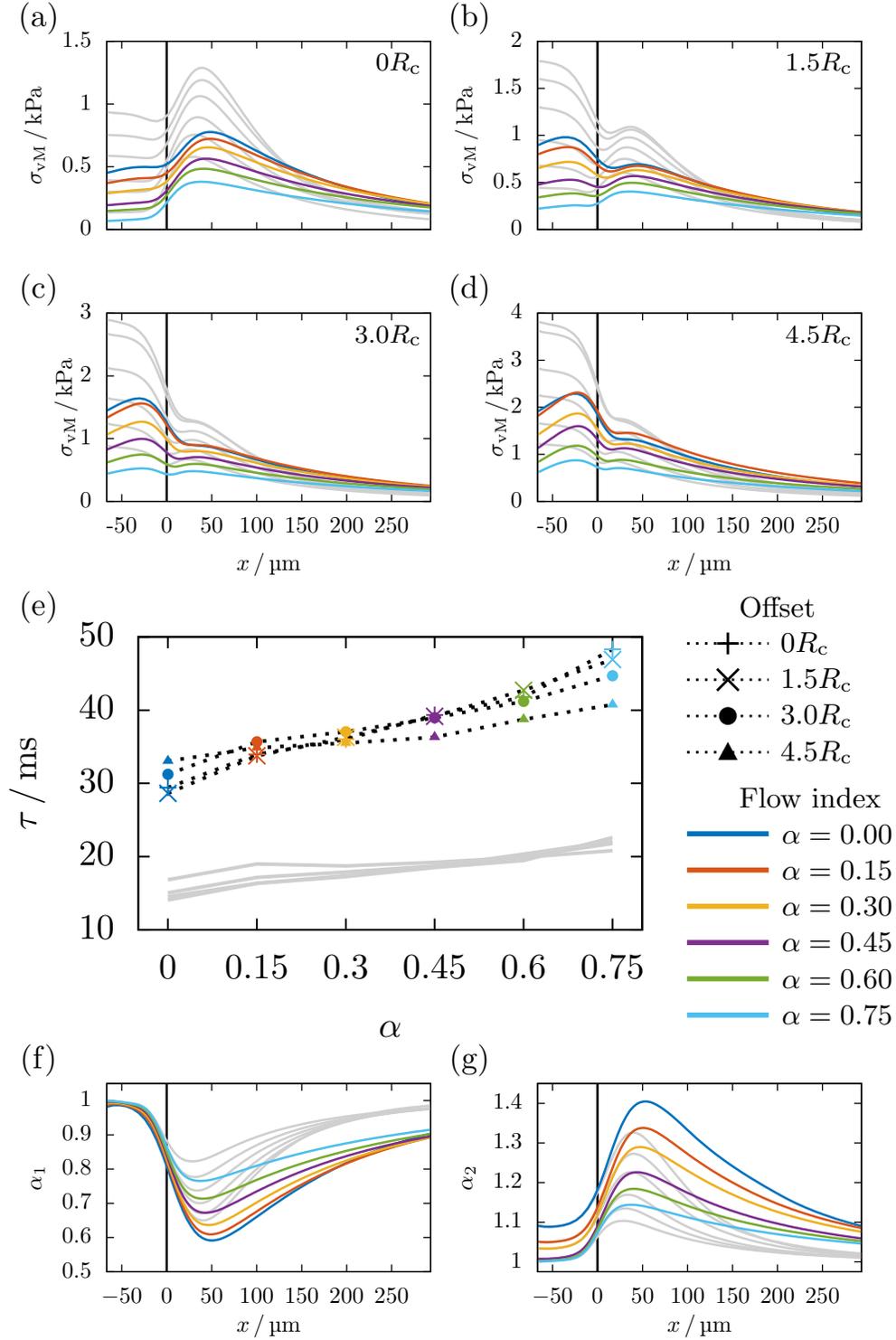

FIG. S-9. Estimated elongational stress at the nozzle exit for a bioink with shear thinning exponent $\alpha = 0.6$ in differently sized nozzles and with (a) variable extrusion speed $u_{\text{avg}}$, or (b) different flow rate $\Omega$ or (c) different printing pressures.

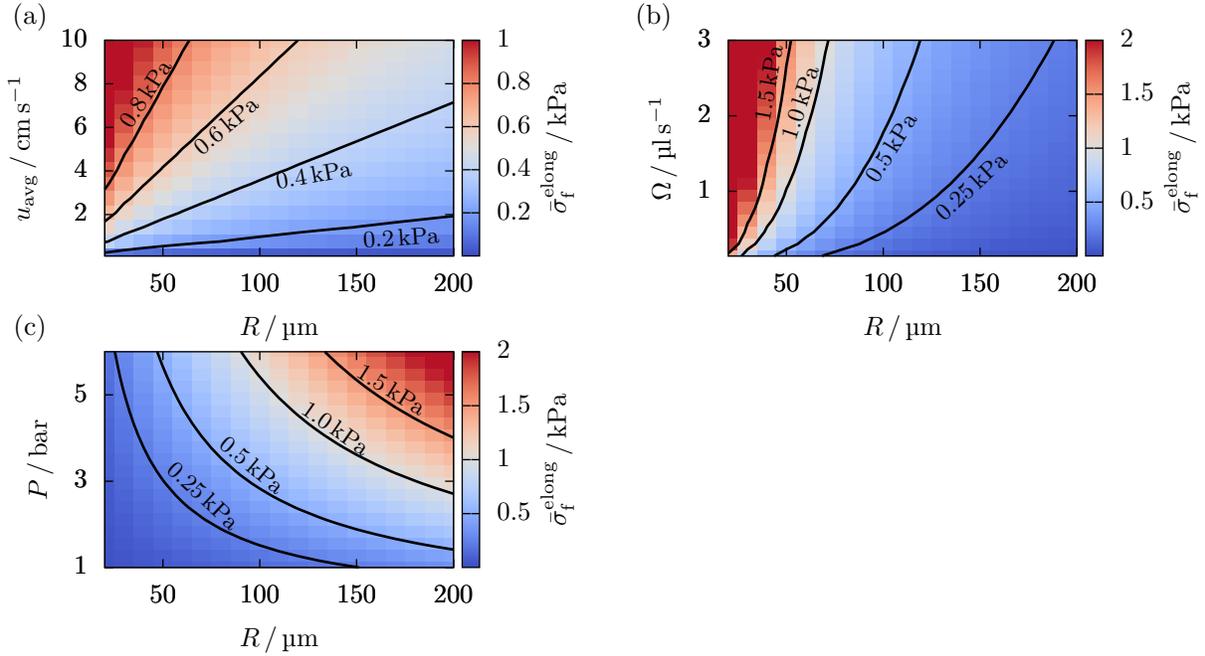

## S-10. ELONGATIONAL FLOW ESTIMATE

To compute the average elongational stress $\bar{\sigma}_f^{\text{elong}}$ acting on cells right at the nozzle exit, we start by calculating the maximum and average flow velocity of our bioink using our tool from [5], as in figure 1(b). From equations (15) and (16) we then obtain $\bar{\sigma}_f^{\text{elong}}$ as a function of the average extrusion velocity $u_{\text{avg}}$ and the nozzle radius $R$. The result is shown in figure S-9, using a the same parameter space as in figure 13 of the manuscript.


\end{document}